\documentclass[Afour,sageh,times]{sagej}

\usepackage{moreverb,url}

\usepackage[colorlinks,bookmarksopen,bookmarksnumbered,citecolor=red,urlcolor=red]{hyperref}
\usepackage{enumitem}
\usepackage{graphicx}
\usepackage{listings}
\usepackage{subfigure}
\usepackage{float}
\usepackage{tabularx}

\lstdefinelanguage{JuliaHPC}{
  sensitive=true,
  morecomment=[l]{\#},
  morestring=[b]",
  alsoletter=@:_,                % <-- make @, :, _ part of words
  morekeywords={
    ForOp,JLIRValue,Vector,Pair,JLIRRegion,function,end,if,else,elseif,while,for,break,continue,return,
    struct,mutable,begin,do,try,catch,finally,import,export,using,
    const,let,in,out,where,abstract,type,module,quote,macro,@inbounds,
    ka,jacc,JACC,Julius,dmem,ones,zeros,arange,linspace,geomspace,
    rand,fill,@jtask,@jparallel_for,@jparallel_reduce,submit,depends,
    kernel,dependencies,Julius.flush,args,softmax_julia,
    CUDA, AMDGPU, Threads, maximum, exp, cuDNN, cudnnTensorDescriptor, cudnnDataType, cudnnSoftmaxForward, cudnnSetTensor4dDescriptor
  },
  morekeywords=[2]{task:,graph:,host:},
  keywordstyle=\color{blue}\bfseries,
  keywordstyle=[2]\color{teal}\bfseries,
}

\usepackage{xcolor}
\usepackage{todonotes}

\def\tabref#1{Table~\ref{#1}}
\def\eqref#1{Equation~\ref{#1}}
\def\lstref#1{Listing~\ref{#1}}

\usepackage{caption} % Required for \captionof

\newcommand\BibTeX{{\rmfamily B\kern-.05em \textsc{i\kern-.025em b}\kern-.08em
T\kern-.1667em\lower.7ex\hbox{E}\kern-.125emX}}

\def\volumeyear{2016}

\begin{document}

\runninghead{Smith and Wittkopf}

\title{JLIR: A Julia-Native MLIR-Inspired Intermediate Representation with 
       Automatic JACC Kernel Extraction}

\author{Narasinga Rao Miniskar\affilnum{1}, Seyong Lee\affilnum{1}, Keita Teranishi\affilnum{1}, Jeffrey S. Vetter\affilnum{1}}

\affiliation{\affilnum{1}Oak Ridge National Laboratory, Oak Ridge, 37830, USA}
%\affilnum{2}SAGE Publications Ltd, UK}

%\corrauth{Alistair Smith, Sunrise Setting Ltd
%Brixham Laboratory,
%Freshwater Quarry,
%Brixham, Devon,
%TQ5~8BA, UK.}

\email{miniskarnr@ornl.gov}

\begin{abstract}
The Multi-Level Intermediate Representation (MLIR) has made reusable compiler infrastructure practical for domain-specific computation. 
%, but extending it still requires substantial compiler-engineering expertise, including C++ and TableGen. 
However, MLIR's strong compile-time type requirements and low-level (C++) extension model can be a poor match for high-level, dynamically specialized languages such as Julia. 
MLIR has several drawbacks for dynamic programming languages in terms of the type system and level of abstraction. It is thus extremely challenging for non-compiler/scientific computing users to introduce new programming abstractions and express algorithm implementations in a form that remains both natural and optimizable. As a result, library interfaces for linear algebra, mesh processing, partial differential equations, and related domains often sit outside the compiler optimization path.
We present \textbf{JLIR} (\textbf{J}ulia-native \textbf{L}evel \textbf{I}ntermediate \textbf{R}epresentation), a Julia-native intermediate representation framework that brings the main benefits of MLIR-style multi-level, dialect-oriented compilation into the Julia ecosystem while remaining usable as ordinary Julia code. JLIR represents Julia programs before low-level lowering, supports extensible operations and transformation passes through Julia's language mechanisms, and allows partially typed programs to remain transformable until concrete types are known. The framework includes built-in dialects for arithmetic, control flow, functions, structured loops, and memory operations, and it also includes a lightweight mechanism for adding new domain operations without modifying the core system. To demonstrate JLIR's capabilities, we applied it to automatic Julia for Accelerators (JACC) kernel generation. JLIR detects parallel and reduction loop patterns and lowers them to JACC kernels, producing portable CPU and GPU code from serial Julia programs without user annotations. Across matrix multiplication, a two-dimensional Jacobi stencil, Black--Scholes option pricing, and transformer model kernels on an NVIDIA A100 GPU, generated kernels achieved up to \textbf{3{,}023~GB/s} effective bandwidth, \textbf{96\%} of a hand-written Julia accelerator baseline, and an \textbf{85$\times$} speedup over serial Julia, with transformation overhead below 1.5~ms. These results show that a dynamic language can host practical, extensible compiler infrastructure while preserving a high-level programming model.

\end{abstract}


\maketitle

\begingroup
\renewcommand{\thefootnote}{}
\footnotetext{\textbf{Notice:} This manuscript has been authored by UT-Battelle, LLC, under contract DE-AC05-00OR22725 with the US Department of Energy (DOE). The US government retains and the publisher, by accepting the article for publication, acknowledges that the US government retains a nonexclusive, paid-up, irrevocable, worldwide license to publish or reproduce the published form of this manuscript, or allow others to do so, for US government purposes. DOE will provide public access to these results of federally sponsored research in accordance with the DOE Public Access Plan ( https://www.energy.gov/doe-public-access-plan ). This work was supported through ``Competitive Portfolios'' DE-FOA-0003264, under award number DE-SC0025645 and FWP ERKJ452.}
\addtocounter{footnote}{-1}
\endgroup

\section{Introduction}
\label{sec:intro}

Modern scientific computing is at the crossroads of two major developments: the rapid expansion of heterogeneous accelerator hardware---including GPUs, field-programmable gate arrays (FPGAs), and specialized machine learning (ML) chips---and the swift uptake of Julia~\cite{julia} as a premier language for numerical and high‑performance computing. Julia provides an attractive blend of interactive productivity and compiled performance, which is realized via LLVM‑based just‑in‑time (JIT) compilation coupled with multiple dispatch. Nevertheless, the compiler infrastructure that supports performance‑portable GPU programming, frameworks such as LLVM~\cite{llvm} and MLIR~\cite{mlir}, is implemented in C++ and accessed through foreign language interfaces that fundamentally clash with Julia's package‑centric, read-eval-print loop (REPL)--first development paradigm.

The central observation motivating this work is that Julia possesses a unique set of language properties that make it an \emph{ideal host} for native Multi-Level Intermediate Representation (MLIR)--style compiler infrastructure:

\begin{description}[leftmargin=0pt, itemsep=2pt]
  \item[\textbf{Homoiconicity.}]
    Julia source code is parsed into \texttt{Expr} objects, which are standard Julia data structures. Consequently, a compiler IR front end can be implemented as a simple Julia function that recursively traverses \texttt{Expr} trees, obviating the need for a dedicated lexer, parser generator, or schema compiler.

  \item[\textbf{Multiple dispatch.}]
    Julia's multiple dispatch mechanism selects method implementations based on the runtime types of all arguments. MLIR's open extensibility model, which permits the addition of new operations and passes without modifying existing code, maps directly onto this mechanism: a new dialect operation is represented by a new Julia type, and a new lowering rule corresponds to a new method definition.

  \item[\textbf{Type-agnostic JIT semantics.}]
    Julia programs are frequently written without explicit type annotations; types are resolved at JIT specialization time. Therefore, an IR for Julia must be capable of representing and transforming code before complete type information is available. This is a constraint that MLIR's strictly typed IR cannot easily satisfy without intrusive analysis.

  \item[\textbf{Meta-programming layer.}]
    Julia's macro system and generated functions operate prior to the language's own LLVM lowering, thereby providing IR transformations with access to a high‑level, semantics‑rich representation that remains hidden from downstream LLVM IR passes.

  \item[\textbf{Package ecosystem.}]
    Julia packages can be installed with a single command and are version‑locked through a manifest; no separate build systems such as CMake or Ninja are required. Consequently, an IR framework delivered as a Julia package becomes instantly available to every Julia user.
\end{description}

\begin{figure}[t]
  \centering
  \includegraphics[width=\columnwidth]{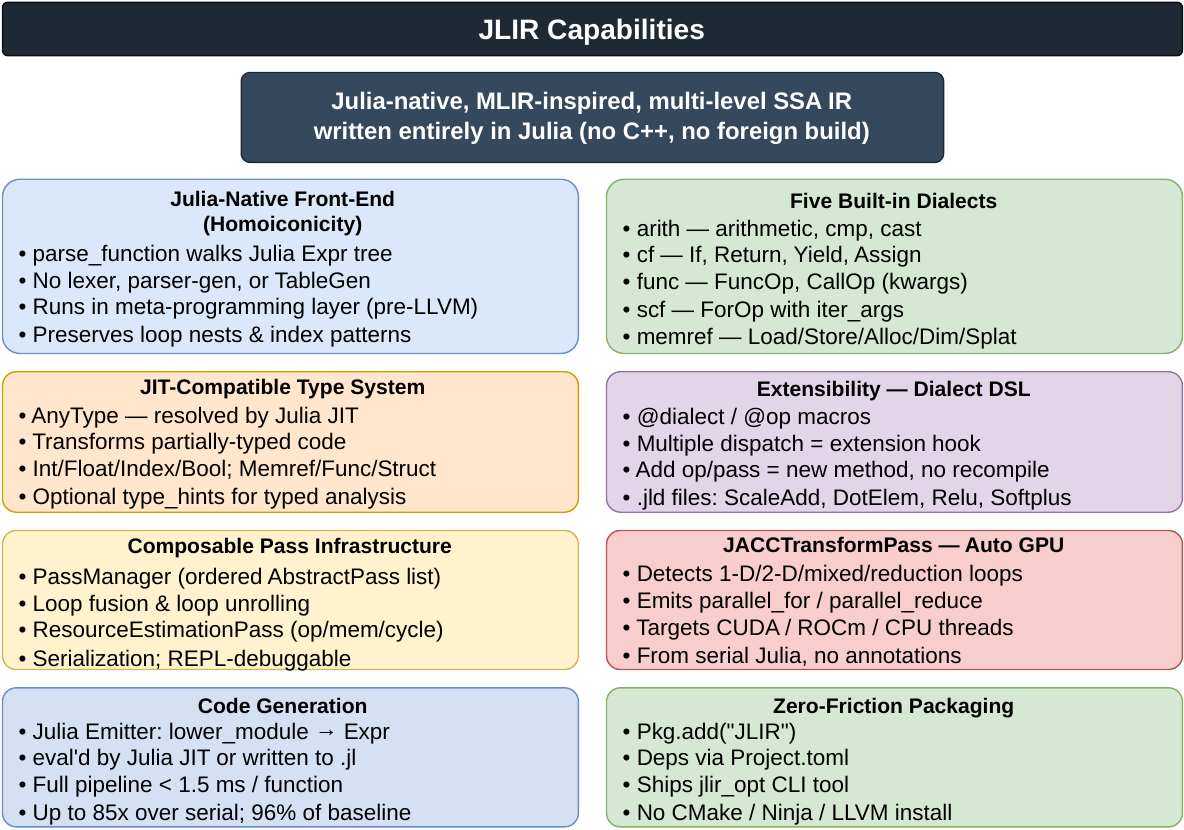}
  \caption{Overview of JLIR's capabilities: a Julia-native front-end, five built-in dialects, a JIT-compatible type system, a composable pass infrastructure, automatic GPU lowering via \texttt{JACCTransformPass}, code generation, and zero-friction packaging.}
  \label{fig:jlir-caps}
\end{figure}

In this paper, we introduce JLIR (\textbf{J}ulia-native \textbf{L}evel \textbf{I}ntermediate \textbf{R}epresentation), a framework that leverages the above properties to provide MLIR-style compiler infrastructure within Julia itself. JLIR is an MLIR‑inspired, multi‑level, static single assignment (SSA)‑based IR framework written entirely in Julia. Figure~\ref{fig:jlir-caps} summarizes its capabilities, showing the Julia-native front end, built-in dialects, type system, composable pass infrastructure, and automatic GPU lowering. Our work's key contributions are as follows.

\begin{itemize}[leftmargin=*]
  \item \textbf{A Julia‑native MLIR abstraction.} JLIR provides a structured SSA IR that features five built‑in dialects (\texttt{arith}, \texttt{scf}, \texttt{cf}, \texttt{memref},  \texttt{func}) that accurately capture Julia's arithmetic, control flow, memory operations, and function definitions---all without any C++ code or foreign build system.

  \item \textbf{A macro‑based Dialect domain-specific language (DSL).} New IR operations can be specified in only a few lines of Julia code by employing the \texttt{@dialect} macro, which automatically generates the boilerplate that would otherwise be produced by MLIR's TableGen‑derived C++. In this setup, Julia's multiple dispatch is the extensibility mechanism: adding an operation or a lowering rule corresponds to defining a new method.

  \item \textbf{A composable pass infrastructure.} JLIR offers a pass manager modeled after MLIR's, including built‑in passes for loop fusion, loop unrolling, resource estimation, and serialization. Because passes are implemented as ordinary Julia types that are dispatched by Julia's runtime, they compose seamlessly without boilerplate and can be inspected or debugged interactively in the Julia REPL.

  \item \textbf{The JACCTransformPass compiler transformation pass as a demonstration.} As a concrete illustration of JLIR's capabilities, we present a pass that automatically identifies 1D, 2D, mixed, and reduction loop patterns within JLIR's SSA representation and rewrites them as Julia for Accelerators (JACC)~\cite{jacc} \texttt{parallel\_for} and \texttt{parallel\_reduce} constructs. This yields GPU‑portable Julia code that is generated entirely automatically from a serial Julia source, a transformation that would ordinarily necessitate a dedicated C++ pass in MLIR.

  \item \textbf{An end-to-end DSL extensibility demonstration.} We define four custom arithmetic operations (\texttt{ScaleAdd}, \texttt{DotElem}, \texttt{Relu}, \texttt{Softplus}) in a concise \texttt{.jld} dialect file and implement user-defined fusion passes that recognize and replace SSA arithmetic sub-expressions. The fused IR is printed in JLIR text format, its arithmetic cost is captured by the ResourceEstimationPass, and it is lowered to executable Julia. Correctness is then verified against reference implementations, and no changes to JLIR's built-in passes or dialect definitions are required.
\end{itemize}

\begin{figure*}[t]
  \centering
  \includegraphics[width=2\columnwidth]{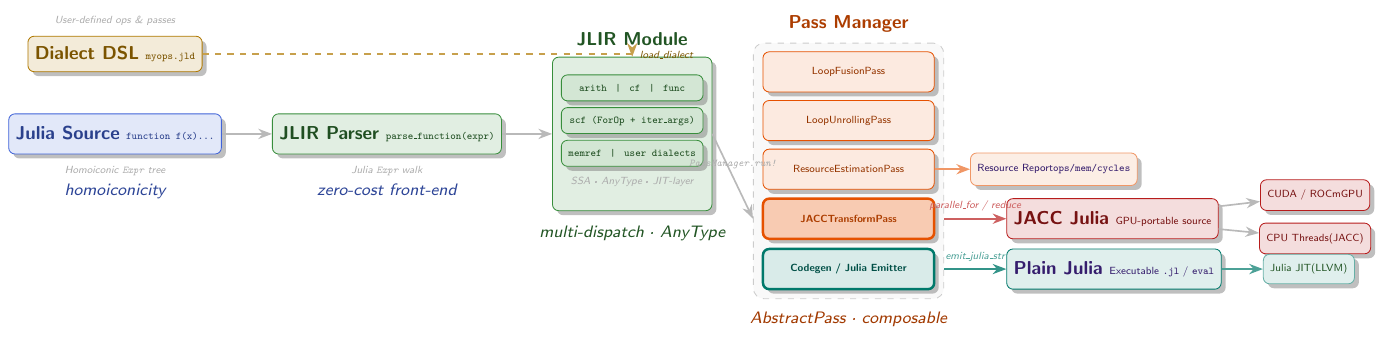}
  \caption{JLIR compilation pipeline. A serial Julia function (\texttt{Expr} tree) is parsed into SSA-form JLIR across five dialects; user-defined operations may be injected via a \texttt{.jld} dialect file (dashed arrow). A configurable pass pipeline then applies transformations: \texttt{JACCTransformPass} emits GPU-portable JACC Julia (targeting CUDA/ROCm/CPU threads), and the \texttt{Codegen / Julia Emitter} pass lowers IR to plain executable Julia evaluated by Julia's JIT. The \texttt{ResourceEstimationPass} produces an operation/memory/cycle report as a side output.}
  \label{fig:arch}
\end{figure*}

Figure~\ref{fig:arch} depicts JLIR's overall compilation pipeline. A Julia source function is parsed into an SSA-form JLIR module spanning five dialects; user-defined dialect operations can be injected via a \texttt{.jld} dialect file (dashed path). The module is then routed through a configurable pass pipeline: \texttt{JACCTransformPass} extracts parallel loops and emits GPU-portable JACC Julia targeting CUDA, ROCm, or CPU threads, while the \texttt{Codegen / Julia Emitter} pass lowers the remaining IR to plain executable Julia evaluated directly by Julia's JIT compiler. For the benchmarks presented in this paper, the complete pipeline completed in less than 1.5~ms, which demonstrates its practicality as a per‑function compilation step.
We benchmarked JLIR on four representative HPC workloads: general matrix--matrix multiplication (GEMM), a 2D Jacobi stencil, Black--Scholes option pricing, and LLaMA‑3 transformer kernels. On an NVIDIA A100 GPU, the automatically generated JACC kernels reached 96\% of a hand‑written Julia+CUDA baseline for Black--Scholes, achieved 3{,}023~GB/s effective bandwidth for the Jacobi stencil ($1.56\times$ the physical DRAM peak), delivered an $85\times$ speedup over serial Julia for the LLaMA-3 \texttt{matmul\_vec} kernel, and reached 87\% of a hand-written Julia+CUDA reference for an untiled DGEMM implementation. Small kernels (dim=4096 LLaMA-3 reductions) fell below the kernel launch amortization threshold; in Section~\ref{sec:results} we discuss this limitation and its implications for automatic lowering. 
%\Seyong{Section numbers are missing here and later}

The remainder of this paper is organized as follows. Section~\ref{sec:background} reviews LLVM, MLIR, Julia, and JACC. Section~\ref{sec:design} presents JLIR's architecture and explains how MLIR's abstractions are mapped onto Julia's language model. Section~\ref{sec:jacc-pass} describes the pass infrastructure in detail, focusing specifically on the automatic GPU kernel extraction algorithm. Section~\ref{sec:benchmarks} introduces the benchmark suite. Section~\ref{sec:results} reports performance results. Section~\ref{sec:relatedwork} surveys related work, and Section~\ref{sec:conclusion} provides conclusions and discusses future work.

\section{Background}
\label{sec:background}

\subsection{LLVM}

The Low Level Virtual Machine (LLVM)~\cite{llvm} is the preeminent open‑source compiler infrastructure employed in systems and scientific computing.  Its core defines a typed, three‑address, SSA~\cite{ssa} IR that functions as a common language between language front ends and machine backends.  A program expressed in LLVM IR is structured into modules, which contain functions, and each function is composed of basic blocks of instructions.  The SSA property ensures that each variable is assigned exactly once and that every use is dominated by its definition. This permits a wide spectrum of classical optimizations, such as constant propagation, dead code elimination, common subexpression elimination, and loop‑invariant code motion, to be formulated as clear, compositional data flow analyses.

For LLVM, the principal advantage lies in its decoupling of language semantics from target architecture. Front ends for a diverse array of languages---including C, Rust, Swift, and Julia---lower programs into LLVM IR, after which a unified optimization pipeline and a suite of backends generate native binaries for targets such as x86, ARM, RISC‑V, NVPTX (NVIDIA GPU), and AMDGPU.  Nevertheless, LLVM IR constitutes a \emph{single‑level} representation that is tuned for machine code emission. Consequently, high‑level program structure (e.g., loop nests, tensor shapes, and memory access patterns) is largely erased by the time the code reaches LLVM IR. This erasure impedes the application of architecture‑aware, domain‑specific transformations such as loop tiling for cache locality, parallel loop extraction for GPU execution, or reduction detection, all of which are essential for contemporary HPC workloads.

\subsection{MLIR}

MLIR~\cite{mlir}, originating at Google and later integrated into the LLVM ecosystem, mitigates the aforementioned issue via a \emph{dialect} architecture. Instead of imposing a fixed IR, MLIR supplies a minimal shared foundation---modules, regions, blocks, operations, values, attributes, and types---and permits an arbitrary number of named \emph{dialects} to enrich this foundation with domain‑specific operation types. The built‑in dialects provided by MLIR include \texttt{arith} for scalar arithmetic, \texttt{cf} for unstructured control flow, \texttt{scf} for structured loops and conditionals, \texttt{memref} for typed memory buffers, \texttt{func} for function definitions and calls, \texttt{linalg} for named tensor contractions, and \texttt{gpu} for device‑side kernel launch.

Dialects are specified via \emph{TableGen}, a domain‑specific schema language that produces the requisite C++ boilerplate for operation structs, verifiers, and printers. MLIR programs can combine dialects freely, and \emph{passes} gradually lower high‑level dialects into more concrete ones until the entire module is represented solely in dialects amenable to translation into LLVM IR. This \emph{progressive lowering} paradigm ensures that each pass remains straightforward and composable, and the \emph{pattern rewriting} infrastructure allows passes to be described declaratively as rewrite rules applied to the IR.

Although MLIR has achieved swift uptake in ML compilers (such as TensorFlow, IREE, and ONNX‑MLIR), quantum computing toolchains, and high-performance computing (HPC) middleware, its reliance on a C++ toolchain and the TableGen compiler decreases accessibility for communities that primarily employ higher‑level languages like Python or Julia.

\subsection{Julia}

Julia~\cite{julia} is a high‑level, high‑performance programming language tailored for numerical and scientific computation. Its hallmark design fuses dynamic typing with LLVM‑based JIT compilation. Method specialization via multiple dispatch produces type‑specific LLVM IR at first execution, enabling performance that rivals statically compiled Fortran or C while maintaining an interactive, expressive syntax like that of Python or MATLAB.

Julia’s compilation pipeline traverses several IR strata. The developer writes Julia source code; the Julia parser constructs an \texttt{Expr} tree, which is a nested array of symbols and literals that is itself a Julia data structure. The front end lowers this tree into a typed \emph{Julia IR}, comparable to a simplified SSA form featuring slots and phi nodes. Subsequent stages translate the Julia IR into LLVM IR, and LLVM’s optimization and code generation passes then emit native machine code. This hierarchical design ensures that Julia’s \texttt{Expr} trees are first‑class, accessible, and manipulable from within Julia.  JLIR exploits the latter capability directly.

Julia’s macro system enables arbitrary compile‑time transformations on \texttt{Expr} trees. Although these macros are potent, they execute prior to the availability of type information, which constrains their capacity to perform type‑driven optimizations or to identify parallelism. In contrast, JLIR operates on the abstract syntax tree (AST) yet builds an enriched SSA IR that retains the structural information required for sophisticated transformations.

\subsection{JACC}
JACC~\cite{jacc} is a Julia package that furnishes a single‑source portability framework for parallel computation. Its two principal entry points are \texttt{JACC.parallel\_for}, which distributes a user‑supplied kernel across an index space, and \texttt{JACC.parallel\_reduce}, which executes a parallel reduction. JACC determines the execution backend at load time according to a configuration preference. On systems equipped with a CUDA‑capable GPU, the default backend employs \texttt{CuArray} and CUDA kernels via CUDA.jl~\cite{cuda_jl}; in the absence of a GPU, it defaults to Julia’s multithreaded \texttt{Threads.@threads}. Support for AMD GPUs via HIP is also available.

The JACC API bears strong resemblance to the Kokkos and RAJA portability layers~\cite{kokkos,kokkos3,raja}, yet it is implemented entirely in Julia and exploits Julia's multiple dispatch to select the appropriate backend during compilation. Concretely, \texttt{JACC.parallel\_for(N, f, args...)} launches \texttt{f(i, args...)} for every integer \texttt{i} in \texttt{1:N} on the active backend. Two‑dimensional dispatch is achieved with a tuple \texttt{parallel\_for((M,N), f, args...)}; this invokes \texttt{f(j, i, args...)} for all \texttt{(j,i)} in $\{1\ldots M\}\times\{1\ldots N\}$. Reductions are performed via \texttt{JACC.parallel\_reduce(f, N, args...; op=+, init=0.0)}. JLIR’s \texttt{JACCTransformPass} targets precisely those two primitives to establish a natural synergy between JLIR and JACC for automated GPU programming.

\section{JLIR Design}
\label{sec:design}

\subsection{Mapping MLIR Abstractions onto Julia's Language Model}
\label{sec:julia_mapping}

JLIR keeps MLIR's core abstractions but implements them in terms of Julia's programming model rather than MLIR's C++ infrastructure. Table~\ref{tab:mapping} shows the corresponding JLIR construct for each MLIR concept, and the following subsections explain the main mappings in more detail.

\begin{table*}[h]
\centering
\caption{Mapping of MLIR concepts to their Julia-native JLIR equivalents.}
\label{tab:mapping}
\begin{tabular}{lll}
\toprule
\textbf{MLIR concept} & \textbf{MLIR mechanism} & \textbf{JLIR (Julia-native)} \\
\midrule
Operation type     & C\texttt{++} struct + TableGen     & Julia \texttt{struct} + \texttt{@op} DSL \\
Dialect definition & TableGen \texttt{.td} file         & \texttt{@dialect} macro \\
Op dispatch        & Virtual method table               & Julia multiple dispatch \\
Pass interface     & C\texttt{++} CRTP$^{\dagger}$ abstract base class & Julia \texttt{AbstractPass} + dispatch \\
Front-end parser   & Custom MLIR parser / C\texttt{++}  & \texttt{Meta.parse} + \texttt{Expr} walk \\
Type system        & Strictly typed at IR build time    & \texttt{AnyType} for JIT-resolved types \\
Extensibility      & New \texttt{.td} + recompile       & New method definitions \\
Installation       & CMake + Ninja + LLVM toolchain     & \texttt{Pkg.add("JLIR")} \\
\bottomrule
\multicolumn{3}{l}{$^{\dagger}$CRTP: Curiously Recurring Template Pattern.} \\
\end{tabular}
\end{table*}

\paragraph{Homoiconicity as a zero-cost front end.}
Within MLIR, the integration of a new source language typically demands the creation of a dedicated C\texttt{++} parser or front end; examples include the Flang Fortran front end and the Torch-MLIR Python front end. Julia’s homoiconic property removes this requirement entirely for Julia programs because the runtime already provides a fully parsed abstract syntax tree as \texttt{Expr} objects, which are ordinary Julia values with subtrees that can be accessed via field lookup. Consequently, the entire front end of JLIR consists of the single function \texttt{parse\_function(expr::Expr)}, which recursively traverses the \texttt{Expr} tree and emits JLIR operations. No tokenizer, grammar specification, or generated code is necessary.

\paragraph{Multiple dispatch as the extensibility mechanism.}
MLIR’s extensibility architecture relies on C\texttt{++} virtual dispatch, whereby each \texttt{Operation} subclass overrides methods such as \texttt{verify()}, \texttt{print()}, \texttt{fold()}, and so on. Julia replaces this mechanism with multiple dispatch, providing equivalent flexibility without a class hierarchy. For instance, adding a new dialect operation \texttt{MyOp} in JLIR requires only the definition of a Julia \texttt{struct MyOp} and the implementation of three associated methods:
\begin{lstlisting}
JLIR.op_results(op::MyOp)   = [op.result]
JLIR.op_operands(op::MyOp)  = [op.lhs, op.rhs]
JLIR.lower_op!(ctx, op::MyOp) =
    push!(ctx.stmts, :($(sym(ctx,op.result)) =
                        $(sym(ctx,op.lhs)) + $(sym(ctx,op.rhs))))
\end{lstlisting}
No existing code needs to be altered, and the framework does not require recompilation. Julia’s method specialization ensures that these dispatch calls achieve performance that is on par with C\texttt{++} virtual calls when the types involved are known at a given call site.

\paragraph{AnyType and JIT‑compatible IR.}
MLIR enforces that every SSA value carries a concrete type at IR construction time. Although this is natural for statically typed source languages, it introduces friction for Julia, in which type annotations are optional and concrete types are typically determined only after JIT specialization. JLIR addresses this by introducing \texttt{AnyType}: an SSA value of \texttt{AnyType} signals that the actual type will be resolved by Julia’s JIT. The JLIR pass infrastructure operates correctly with \texttt{AnyType} values because Julia’s own runtime inserts the appropriate type checks and specializations when the generated code is executed. This feature allows JLIR to transform programs without full type information, mirroring the capabilities of Julia’s macro system. When explicit type hints are supplied via a \texttt{type\_hints} dictionary argument to the parser, JLIR employs concrete \texttt{IntType} or \texttt{FloatType} values, enabling limited type‑directed analysis.

\paragraph{Meta‑programming layer integration.}
Julia’s compilation pipeline performs \texttt{Expr} macro expansion and generated function handling prior to type inference and LLVM lowering. JLIR operates within this same meta‑programming layer: a call such as \texttt{parse\_function(:(function f(x)...\,end))} is executed during Julia’s meta‑programming phase, not after LLVM IR has been produced. As a result, JLIR has access to the complete syntactic structure of the program (e.g., loop nests, array index patterns, conditional branches) before any information is lost during lowering. In contrast, an LLVM pass operates on already lowered, type‑erased IR where loop nests may appear as pointer arithmetic and array accesses as raw \texttt{getelementptr} instructions.

\paragraph{Zero‑friction packaging.}
Every design decision in JLIR is fully compatible with Julia’s standard package ecosystem. The framework is distributed as a single registered Julia package; its dependencies (CUDA.jl, JACC.jl, SpecialFunctions.jl) are declared in \texttt{Project.toml} and resolved automatically by the Julia package manager. A user installs JLIR with \texttt{Pkg.add("JLIR")} and immediately gains access to the complete pass infrastructure, the dialect library, and the \texttt{jlir\_opt} command‑line tool.  It does not require CMake, Ninja, or a separate LLVM installation.

\subsection{IR Data Structures}

JLIR’s intermediate representation is designed to mirror the container hierarchy of MLIR while providing an abstraction layer that is specifically suited to the representation of Julia programs.

\paragraph{JLIRValue.}
The elementary entity in this IR is an SSA value, which is identified by a globally unique integer ID.  Each value can optionally carry the name of the source‑level variable, a \texttt{JLIRType} (which can be \texttt{AnyType}), and an optional compile‑time constant.  JLIR embeds constant values directly into the IR to enable constant folding during construction and facilitates the early detection of loop bounds, thereby reducing runtime overhead.

\paragraph{JLIROp.}
Every computation is expressed as an \emph{operation} (\texttt{JLIROp}), an abstract Julia type that defines two dispatch methods: \texttt{op\_results} for the SSA values it produces and \texttt{op\_operands} for the SSA values it consumes.  Structured operations encapsulate one or more \texttt{JLIRRegion}s, and these regions contain sequential lists of \texttt{JLIRBlock}s that comprise further operations.  This design preserves the structural clarity of MLIR while allowing Julia’s rich control flow constructs to be represented naturally.

\paragraph{JLIRModule.}
The top-level container holds an ordered list of operations (typically \texttt{FuncOp}
and \texttt{StructDefOp}) and a module-level attribute dictionary.  A
\texttt{Builder} maintains the running SSA counter, the current block pointer, and
a lexical scope environment that maps Julia symbol names to their current SSA bindings.

\subsection{Type System}

The \texttt{Types} module defines a type hierarchy that extends from the abstract \texttt{JLIRType}.  Primitive types include \texttt{IntType(width, signed)}, \texttt{FloatType(width)}, \texttt{IndexType} (intended for loop induction variables), \texttt{BoolType}, and \texttt{AnyType}.  Composite types include \texttt{MemrefType(element, shape)}, \texttt{FuncType}, and \texttt{StructType}.  Singleton constants (\texttt{i8}, \texttt{f64}, etc.) are exported at the module level and thus provide convenient access to frequently used types.

\subsection{Dialects}

JLIR incorporates five core dialects, each defined as a sub‑module within \texttt{JLIR.Dialects}.

\paragraph{Arith.}
This dialect implements scalar arithmetic and logical operations.  The set of opcodes includes \texttt{ConstantOp}, \texttt{BinOp}, \texttt{UnaryOp}, \texttt{CmpOp} (supporting ten integer and fourteen floating point predicates), and \texttt{CastOp} (providing ten cast varieties).  The design preserves MLIR’s efficient handling of elementary operations and exposes a Julia‑friendly API.

\paragraph{CF.}
Control flow primitives that do not own regions are provided by the CF dialect.  It defines \texttt{IfOp} (with separate then/else regions), \texttt{ReturnOp}, \texttt{YieldOp}, and \texttt{AssignOp}.  The \texttt{AssignOp} is particularly noteworthy because it models Julia’s mutable variable rebinding by generating a new SSA value that shares the same source‑level name, thereby preserving Julia’s semantics of variable mutation within an SSA framework.

\paragraph{Func.}
The \texttt{FuncOp} encapsulates a Julia function together with its body region.  \texttt{CallOp} supports invocation of both named and first‑class callee values; keyword arguments are encoded in an \texttt{attributes} field.  This mechanism is leveraged by \texttt{JACCTransformPass} to inject keyword arguments such as \texttt{op=+, init=0.0} into \texttt{parallel\_reduce} calls, thereby enabling fine‑grained control over parallel reduction semantics.

\paragraph{SCF (Structured Control Flow).}
Structured control flow constructs are implemented in the SCF dialect.  \texttt{ForOp} models a Julia \texttt{for i in lb:step:ub} loop; its \texttt{iter\_args} attribute stores pairs of the form (initial‑value, loop‑variable) that capture loop‑carried values.  The content of \texttt{iter\_args} is a critical indicator used by  \texttt{JACCTransformPass} to distinguish embarrassingly parallel loops (empty \texttt{iter\_args}) from reduction loops (exactly one entry), thus enabling targeted optimizations.

\paragraph{Memref.}
Memory reference operations are grouped in the Memref dialect.  \texttt{LoadOp} and \texttt{StoreOp} represent indexed array reads and writes, whereas \texttt{AllocOp}, \texttt{DimOp}, and \texttt{SplatOp} provide allocation and shape query primitives.  This suite of operations mirrors MLIR’s Memref dialect to ensure efficient memory manipulation while staying compatible with Julia’s array abstractions.

\subsection{Dialect DSL}

The \texttt{@dialect} macro automates the generation of all boilerplate required for a new dialect operation based on a succinct inline specification as shown in \lstref{lst:myops}.

\begin{figure}[t]
\centering
\begin{lstlisting}[label={lst:myops}, caption={User-defined dialect example}]
@dialect "myops" MyDialect begin
  @op "myops.scale_add" ScaleAddOp begin
    summary  = "Fused multiply-add: result = lhs + scale * rhs"
    operands = [lhs, rhs, scale]
    results  = [result]
    lower    = :(lhs + scale * rhs)
  end
end
\end{lstlisting}
\end{figure}

The macro expands into a \texttt{ScaleAddOp} struct, its \texttt{op\_results} and \texttt{op\_operands} methods (dispatched by Julia’s runtime), a \texttt{Base.show} implementation that conforms to JLIR’s textual format, and a \texttt{lower\_op!} method for the code generator.  The macro itself is implemented using Julia’s \texttt{Expr} manipulation facilities---the same mechanism that underpins JLIR’s parser.

\subsection{Parser}

The parser, implemented as \texttt{JLIR.}\texttt{Parser.}\texttt{parse\_function}, consumes a \texttt{:function} \texttt{Expr} and yields a \texttt{FuncOp}.  A \texttt{ParserContext} encapsulates a \texttt{Builder} and optionally carries type hint annotations.  During parsing, the dispatcher examines the \texttt{head} field of each \texttt{Expr}: an \texttt{:=} expression triggers assignment, loop constructs generate \texttt{ForOp} nodes after performing range analysis, and conditional expressions produce \texttt{IfOp} nodes with two regions.  Importantly, a \texttt{for} loop iterating over a range and accumulating into a mutable variable is immediately encoded as a \texttt{ForOp} with a non‑empty \texttt{iter\_args} list. This renders the reduction pattern explicitly visible during IR construction, obviating the need for later analysis.

\subsection{Pass Manager and Code Generator}

The \texttt{PassManager} maintains an ordered sequence of \texttt{AbstractPass} instances.  Each pass consumes a \texttt{JLIRModule} and yields a \texttt{PassResult}.  Because passes are plain Julia types and dispatch is handled via Julia’s multiple dispatch, a user can implement a new pass as a single Julia struct that exposes one method, insert it into the pipeline, and execute it directly from the REPL.  The code generator, implemented in \texttt{JLIR.Codegen.lower\_module}, translates the transformed module into a Julia \texttt{Expr} that can be \texttt{eval}’d or written to a \texttt{.jl} file.

\section{Use Case: JACC Transform Pass}
\label{sec:jacc-pass}

The JLIR framework now incorporates optimization and inspection passes that enable efficient conversion of Julia kernels to JACC.
\texttt{LoopFusionPass} merges consecutive ForOps with identical bounds and no data hazards, reducing loop overhead. LoopUnrollingPass expands loops
with known trip counts, either fully or partially, to expose parallelism. PrinterPass produces an MLIR‑style textual form of the
JLIRModule, while TextParser rebuilds it, allowing round‑trip inspection. ResourceEstimationPass statically counts FLOPs, loads, and
stores to attach memory traffic and compute density attributes.  This section describes the ultimate use case of JLIR in transforming the JLIR code to vendor-agnostic automatic multi-dispatch JACC kernels. 

\texttt{JACCTransformPass} is the core capability of JLIR; it traverses each \texttt{FuncOp} within the module and automatically transforms suitable loop nests into calls to \texttt{JACC.parallel\_for} or \texttt{JACC.parallel\_reduce} kernels, thereby generating GPU‑ready Julia code from plain serial sources without any explicit annotations from the developer.

\subsection{The Central Detectability Predicate}

The entire detection logic hinges on a single structural attribute of a \texttt{ForOp}: its \texttt{iter\_args} field. Within JLIR’s SSA representation, a \texttt{ForOp} is defined in \lstref{lst:forop}.

\begin{figure}[t]
\centering
\begin{minipage}{\linewidth}
\begin{lstlisting}[label={lst:forop}, caption={JLIR's ForOp single structural attribute}]
ForOp(
  induction_var :: JLIRValue,   # loop index i
  lower_bound   :: JLIRValue,
  upper_bound   :: JLIRValue,
  step          :: JLIRValue,
  iter_args     :: Vector{Pair{JLIRValue, JLIRValue}},
  body          :: JLIRRegion
)
\end{lstlisting}
\end{minipage}
\end{figure}

Each element of \texttt{iter\_args} is a pair \texttt{(init\_val, loop\_var)} in which \texttt{init\_val} denotes the value before the loop begins and \texttt{loop\_var} is its in‑loop name. A non‑empty \texttt{iter\_args} set indicates the presence of loop‑carried state, hence a sequential dependency; conversely, an empty \texttt{iter\_args} denotes embarrassingly parallel work.

\begin{center}
\fbox{\parbox{1.0\columnwidth}{
\textbf{Detectability rule:}~~
\texttt{isempty(for\_op.iter\_args)} $\Rightarrow$ \textbf{parallel}
\quad;\quad
\texttt{length(for\_op.iter\_args) == 1} with a matching reduction body
$\Rightarrow$ \textbf{reducible}
}}
\end{center}

This predicate is computed by the function \texttt{\_is\_parallel(for\_op)} in one line of Julia:

\begin{lstlisting}
_is_parallel(f::ForOp) = isempty(f.iter_args)
\end{lstlisting}

The pass conducts a single linear traversal of each function: to ensure that reductions are not mistaken for parallel loops, it first processes potential reduction loops before handling ordinary parallel nests.

\begin{table*}[h]
\centering
\caption{JACC Transformation Pass with four different pattern matches and examples of input Julia, JLIR, and output JACC transformation}
\label{tab:jacc-patterns}
\begin{tabular}{rrrrr}
\toprule
\textbf{Pattern} & \textbf{Kernel} & \textbf{Julia Source} & \textbf{JLIR} & \textbf{Generated JACC} \\
\midrule
1D Parallel-for & rms\_norm\_scale & \lstref{lst:1dparallel-julia} &  \lstref{lst:1dparallel-jlir} &  \lstref{lst:1dparallel-jacc} \\
2D Parallel-for & jacobi2d & \lstref{lst:2dparallel-julia} &  \lstref{lst:2dparallel-jlir} &  \lstref{lst:2dparallel-jacc} \\
2D Parallel-for & gemm & \lstref{lst:2dparallel-mixed-julia} &  \lstref{lst:2dparallel-mixed-jlir} &  \lstref{lst:2dparallel-mixed-jacc} \\
(Mixed 2D+Sequential) & & & & \\
1D Parallel-reduce & rms\_norm\_reduce& \lstref{lst:1dparallel-reduce-julia} &  \lstref{lst:1dparallel-reduce-jlir} &  \lstref{lst:1dparallel-reduce-jacc} \\
\bottomrule
\end{tabular}
\end{table*}

\subsection{Detection of Patterns}
The four different detection patterns used in \texttt{JACCTransformPass} are shown in \tabref{tab:jacc-patterns}.

\paragraph{1D parallel-for detection:} 
During detection, \texttt{\_is\_parallel(for\_op)} evaluates to \texttt{true}, while \texttt{\_inner\_for(for\_op)} returns \texttt{nothing} (no nested loop). This classifies the construct as a 1D parallel pattern.

\paragraph{2D parallel-for detection:}
Here, the outer loop satisfies \texttt{\_is\_parallel} → \texttt{true}. The inner loop is retrieved via \texttt{\_inner\_for} and also passes the \texttt{\_is\_parallel} test. Consequently, the two‑dimensional structure is identified as a 2D parallel pattern.

\paragraph{2D Parallel-for Mixed+sequential detection:} 
The outer loops satisfy \texttt{\_is\_parallel} → \texttt{true}. The innermost loop fails the parallel test because it contains a loop‑carried accumulator. Consequently, the pass recognizes this as a mixed 2D parallel loop where the inner sequential body is preserved within the kernel.

\paragraph{1D parallel-reduce detection via \texttt{\_detect\_reduce}:}
The \texttt{\_detect\_reduce} routine evaluates four sequential conditions (\lstref{alg:detect_reduce}): it first verifies that the loop carries exactly one value, extracts the initial value and loop-carried variable, locates the \texttt{AssignOp} that updates the loop-carried variable, finds the \texttt{BinOp} that generates the assigned value, and finally ensures the operator belongs to the set \{+,*,min,max\}. When these checks succeed, the routine isolates the per‑element operand and returns a tuple containing the initial value, the per‑element contribution, and the reduction operator.
\begin{enumerate}[leftmargin=*, itemsep=1pt]
  \item \texttt{length(iter\_args) == 1} $\to$ satisfied (\texttt{\%ss0, \%ss}).
  \item An \texttt{AssignOp} targeting \texttt{\%ss} exists $\to$ found (\texttt{\%ss = assign \%ss}).
  \item A \texttt{BinOp} produces the assigned value $\to$ \texttt{BinOp(:+)}.
  \item One operand of the \texttt{BinOp} is the accumulator
        (\texttt{\%ss} $\in$ \texttt{acc\_ids}) $\to$ satisfied;
        the other operand \texttt{\%xi2} is the per-element value.
  \item \texttt{reduce\_op = :+} $\in$ \texttt{\{+, *, min, max\}} $\to$ valid.
\end{enumerate}
The function returns \texttt{(init\_val=\%c0, per\_elem=\%xi2, op=:+, ...)}.

\subsection{Free Variable Collection}

After a loop is categorized, the pass must identify all SSA values originating outside the loop body that are referenced within it (e.g., operands to \texttt{LoadOp}, \texttt{StoreOp}, \texttt{CallOp}, and so forth). These values become parameters of the extracted kernel. The \texttt{\_free\_vars} routine achieves this through the following two‑step process.

\begin{enumerate}[leftmargin=*, itemsep=1pt]
  \item Recursively walk all ops in the kernel body and sub-regions,
        collecting every SSA value \emph{ID defined} within (including the
        induction variable and any \texttt{iter\_arg} loop-carried variables).  These
        form the \emph{local definition set} \texttt{local\_defs}.
  \item Walk the ops again, collecting every \emph{operand} whose ID is
        \emph{not} in \texttt{local\_defs}.  These are the free variables,
        returned in first-use order and deduplicated.
\end{enumerate}

The free variables become the trailing parameters of the kernel function
signature, and the same list (in the same order) becomes the trailing arguments
of the \texttt{JACC.parallel\_for} or \texttt{JACC.parallel\_reduce} call.
This guarantees that the call site and the kernel definition always agree on
argument order without any symbol table lookup.

\subsection{Decision Flow}

\lstref{alg:detect_parallel} presents the comprehensive decision logic of \texttt{\_transform\_func!} for each \texttt{ForOp} encountered during a linear scan of a function body.

%\begin{figure*}[h]
\begin{figure}[h]
%\begin{minipage}{0.47\linewidth}
\begin{minipage}{\linewidth}
\begin{lstlisting}[label={alg:detect_parallel}, caption={Algorithm for complete decision flow of \texttt{JACCTransformPass} for each \texttt{ForOp} in a function body.  The scan is linear (one pass); the \texttt{iter\_args} field of each \texttt{ForOp} is the sole structural signal.}]
For each op in function body (linear scan):
  if op is ForOp and NOT parallel:
    rd = _detect_reduce(op)
    if rd != nothing:
      → parallel_reduce(emit kernel + 
                JACC.parallel_reduce)
    else:
      → leave sequential (skip)
  if op is ForOp and IS parallel:
    inner = _inner_for(op)
    if inner == nothing:
      → 1-D parallel_for (body ops → kernel, 
          N → parallel_for(N,...))
    elif _is_parallel(inner):
      → 2-D parallel_for (inner body ops → kernel,
          tuple(N_inner,N_outer) → parallel_for)
    else:  (inner has iter_args → sequential)
      → Mixed parallel_for (entire outer body → kernel,
          N_outer → parallel_for(N_outer,...))
\end{lstlisting}
\end{minipage}
\end{figure}
%\hfill
\begin{figure}[h]
%\begin{minipage}{0.47\linewidth}
\begin{minipage}{\linewidth}
\begin{lstlisting}[label={alg:detect_reduce}, caption={Algorithm for reduction detection predicate.  Checks four structural conditions on the \texttt{ForOp}'s \texttt{iter\_args} and body ops. Both \texttt{init\_val.id} and \texttt{loop\_var.id} are in \texttt{acc\_ids} because Julia semantics reuse the same variable name for successive assignments.}]
_detect_reduce(for_op):
  if |iter_args| != 1:          return None
  (init_val, loop_var) = iter_args[1]
  find AssignOp with target == loop_var
  if not found:                 return None
  find BinOp producing assigned value
  if not found:                 return None
  if BinOp.op not in {+,*,min,max}: return None
  if init_val in BinOp.operands OR loop_var in BinOp.operands:
      per_elem = the other operand
      return (init_val, per_elem, BinOp.op)
  return None
\end{lstlisting}
\end{minipage}
\end{figure}
%\end{figure*}

\subsection{Why \texttt{iter\_args} Is the Right Signal}

The distinction between parallel and sequential loops, as captured by the presence or absence of \texttt{iter\_args}, reflects MLIR’s established convention for loop‑carried values, which JLIR concretizes in its SCF dialect. In MLIR’s \texttt{scf.for}, the \texttt{iter\_args} operands play the same role. JLIR adopts this design and leverages it at a higher level: the JLIR parser---which is implemented as a Julia function that walks \texttt{Expr} trees---can directly inspect the original Julia source to determine whether a variable is updated inside the loop, encoding that knowledge as \texttt{iter\_args} during IR construction. Consequently, the transformation pass can read this pre‑computed structural fact in $O(1)$, enabling the entire detection algorithm to operate as a single linear scan without requiring fixed‑point iteration, alias analysis, or other expensive data flow analyses. This deliberate alignment with MLIR not only guarantees correctness but also yields a highly efficient, lightweight optimization pipeline.

\subsection{Limitation}
A current limitation of the JACCTransformation pass is that it creates JACC arrays for every input and output Julia array, resulting in additional data transfer overhead and reduced performance efficiency. It requires more additional passes such as global scope data flow analysis and reuse of Julia arrays, which is a future work for this paper.
\section{Benchmarks}
\label{sec:benchmarks}

To demonstrate JLIR, we chose four benchmarks that represent complementary computational patterns in order to cover the principal workloads encountered in scientific computing and ML inference.
For each benchmark we preserved a plain serial Julia implementation, referred to as the \emph{source}.  This source is parsed and transformed automatically by the JLIR pipeline without any manual annotations; this feature is a novel contribution.  We evaluated four distinct execution variants: (1) the original Julia serial loop; (2) a hand‑crafted Julia+CUDA kernel written directly using \texttt{CUDA.@cuda} that implements the identical algorithm and serves as a performance baseline; (3) the JACC GPU kernel generated by JLIR; and (4) the identical JLIR‑generated kernel executed on CPU threads via the JACC thread backend.  For GEMM, an additional cuBLAS~\cite{cublas} baseline was included to represent the vendor‑optimized ceiling.

\subsection{Matrix--Matrix Multiplication (GEMM)}

The GEMM benchmark performs the matrix multiplication $C \leftarrow A \cdot B$ on square $M \times M$ matrices using double precision.  The serial implementation consists of a conventional three‑level loop nest: the outer two loops iterate over the output indices $(i,j)$, whereas the innermost $k$ loop accumulates the dot product.  In the JLIR representation, the $k$ loop is encoded as a \texttt{ForOp} with a single \texttt{iter\_arg} that stores the accumulator; the two outer loops are represented as nested \texttt{ForOp}s without \texttt{iter\_args}. \texttt{JACCTransformPass} identifies this structure as a \emph{Mixed 2D + sequential} pattern (see Section \ref{sec:jacc-pass}) and produces a two‑dimensional JACC kernel in which the two outer dimensions are distributed by JACC, whereas the sequential $k$ loop remains within the kernel body.  This automatic classification and kernel generation by \texttt{JACCTransformPass} is an innovative aspect of our approach.  The Julia+CUDA reference employs the identical three‑level loop formulation compiled directly with CUDA.jl’s \texttt{@cuda} kernel launch interface, and an additional cuBLAS baseline is included to represent the vendor‑optimized upper bound.

\subsection{Jacobi 2D Stencil}

The Jacobi stencil implements a single iteration of the five‑point Laplacian average over the interior of a padded grid of size $(M+2)\times(N+2)$ in double precision:
\begin{equation}
  u^{\text{new}}_{i+1,j+1} = \tfrac{1}{4}\bigl(
    u_{i,j+1} + u_{i+2,j+1} + u_{i+1,j} + u_{i+1,j+2}
  \bigr).
\end{equation}
The nested two‑level loop over $(i,j)\in\{1\ldots M\}\times\{1\ldots N\}$ is trivially parallelizable.  \texttt{JACCTransformPass} categorizes this as a \emph{2-D parallel} pattern and emits
\(\texttt{JACC.parallel\_for}((N,M), \texttt{jacobi2d\_kernel}, \\
\text{inp}, \text{out})\).
Within the kernel, both grid indices are supplied, and all offset arithmetic ($i\pm1$, $j\pm1$) is performed internally, with identity constants generated by the JLIR code generator.  The workload is primarily limited by memory bandwidth; we evaluated the effective bandwidth in GB/s by accounting for four loads and one store per output element.  The direct generation of a two‑dimensional parallel kernel from the serial source constitutes an innovative step in our pipeline.

\subsection{Black--Scholes Option Pricing}

The Black--Scholes benchmark~\cite{black_scholes_orig} computes European call and put option prices for $n$ independent contracts in double precision using the closed‑form expressions:
\begin{align}
  C_i &= S_i N(d_1) - K_i e^{-rT} N(d_2), \\
  P_i &= K_i e^{-rT} N(-d_2) - S_i N(-d_1),
\end{align}
where $d_1$ and $d_2$ contain evaluations of $\log$, $\exp$, and $\operatorname{erf}$.  The body of the loop is highly compute‑intensive, and each option is processed independently.  Four scalar parameters---$\sqrt{T}$, $\sigma\sqrt{T}$, $\sigma^2/2$, and $e^{-rT}$---are pre‑computed before entering the loop to avoid redundant transcendental function calls for each thread.  In the JLIR representation, these scalars appear as \texttt{ConstantOp}s or locally defined SSA values outside the \texttt{ForOp}, and  \texttt{JACCTransformPass} properly treats them as free‑variable arguments of the extracted one‑dimensional kernel.  On the GPU, CUDA.jl routes the $\operatorname{erf}$ call to the CUDA math library’s device intrinsic, whereas on the CPU the same functionality is supplied by SpecialFunctions.jl.  Performance is reported in estimated GFLOP/s, with an approximate count of 20 floating‑point operations per option.  The ability to treat transcendental scalars as free variables in the kernel showcases the flexibility of the JLIR representation.

\subsection{LLaMA-3 Inference Kernels}

The LLaMA‑3~\cite{llama3} benchmark invokes three micro‑kernels extracted from the attention and feed‑forward layers of a large language model.  These micro‑kernels comprise (1) token‑embedding lookup, which is implemented as a gather operation over a row‑major embedding table; (2) root mean square (RMS) normalization, which involves a reduction over a hidden dimension vector followed by a scaling pass; and (3) Rotary Position Embedding (RoPE)-NeOX positional encoding, realized as an in‑place rotation of complex pairs in the query and key tensors.  The RMS norm reduction constitutes the sole benchmark that triggers the use of \texttt{parallel\_reduce}: the serial source code contains \(\texttt{ss} += x[i]^2\), which the JLIR parser encodes as a \texttt{ForOp} with a single \texttt{iter\_arg}.  The reduction detector within  \texttt{JACCTransformPass} (Listing~\ref{alg:detect_parallel}) subsequently rewrites this into a \texttt{JACC.parallel\_reduce} invocation.  Both the embedding lookup and RoPE kernels are mapped to one‑dimensional \texttt{parallel\_for} patterns.  The detection and transformation of reductions into \texttt{parallel\_reduce} exemplifies the advanced analytical capabilities of  \texttt{JACCTransformPass}.

\subsection{Experimental Platform}

All experiments were conducted on a server equipped with 128 CPU threads (two AMD EPYC sockets) and an NVIDIA A100~\cite{a100} Peripheral Component Interconnect Express (PCIe) GPU with 80~GB of memory.  The A100 PCIe variant has a physical peak memory bandwidth of approximately 1{,}935~GB/s (the SXM4 variant peaks at ${\approx}2$~TB/s); the peak bandwidth of a single CPU socket is roughly 300~GB/s.  The software environment consists of Julia 1.12.5, CUDA.jl 5.11.0, and JACC.jl 1.1.0.  GPU runtimes were measured using \texttt{CUDA.synchronize} barriers, whereas \texttt{@elapsed} was used for CPU runtimes.  Reported execution times correspond to the \emph{minimum} observed over five repetitions for CPU runs and ten repetitions for GPU runs, after excluding warmup iterations.  The minimum statistic is standard practice for microbenchmarks because it best approximates the hardware-limited cost by suppressing OS noise and thermal throttling. For the Jacobi stencil experiments, \emph{effective} (logical) bandwidth computed as $(4\,\text{loads}+1\,\text{store})\times 8\,\text{B}/t$ and may exceed the physical DRAM peak due to L2 cache hit reuse in the five-point stencil access pattern.

\section{Results}
\label{sec:results}

\subsection{JLIR Pipeline Compilation Overhead}

An important consideration for a source‑level IR framework is the overhead incurred by the transformation pipeline. Whereas LLVM or MLIR manipulate compiled artifacts, JLIR directly operates on Julia \texttt{Expr} trees, a lightweight purely symbolic representation. Table~\ref{tab:compile_overhead} documents the wall‑clock duration required to run the complete JLIR pipeline---which includes parsing, \texttt{JACCTransformPass}, and code emission---for each benchmark source function.

\begin{table}[h]
\centering
\caption{JLIR pipeline compilation overhead per source function. LOC: Lines of Code}
\label{tab:compile_overhead}
\small
\setlength{\tabcolsep}{4pt} % reduce column padding
\begin{tabularx}{\columnwidth}{l X r}
\toprule
\textbf{Benchmark} & \textbf{Source LOC} & \textbf{Pipeline Time} \\
\midrule
GEMM                & 10 & $<0.4$~ms \\
Jacobi 2D stencil  & 8  & $<0.3$~ms \\
Black--Scholes      & 12 & $<0.5$~ms \\
LLaMA-3 (3 kernels) & 52 & $<1.2$~ms \\
\bottomrule
\end{tabularx}
\end{table}

Every benchmark finishes the JLIR transformation in less than 1.5~ms.  This overhead is fully amortized by Julia’s intrinsic JIT compilation, which ordinarily consumes tens to hundreds of milliseconds on the first invocation of a new method specialization.  Consequently, the JLIR pipeline introduces a negligible additional end‑to‑end cost relative to the overall application runtime.

\subsection{GEMM Performance}
\label{sec:results:gemm}

Table~\ref{tab:gemm} presents DGEMM ($C \leftarrow A \cdot B$, \texttt{Float64}) throughput over matrix sizes $N \in \{256,\ldots,16384\}$ for four backends: JACC CPU (threads), cuBLAS, JACC GPU, and a hand‑written Julia+CUDA kernel that implements the identical three‑level untiled loop directly via \texttt{@cuda}.  The Julia+CUDA kernel serves as the ground truth reference for the algorithmic cost, isolating any overhead introduced by the JLIR code generation pipeline.  All seven sizes were benchmarked for Julia+CUDA, including $N=8192$ and $N=16384$, for which each kernel invocation took 4--35~s.

\begin{table}[h]
\centering
\caption{DGEMM throughput (NVIDIA A100 PCIe, \texttt{Float64}).
         JACC CPU is omitted (---) for $N>2048$.
         JACC GPU and Julia+CUDA both use the same untiled $O(N^3)$ loop; cuBLAS uses double-precision (FP64) tensor cores with shared-memory tiling.
         All GPU values in TF/s; JACC CPU in GF/s.}
\label{tab:gemm}
\small
\setlength{\tabcolsep}{4pt}
\begin{tabularx}{\columnwidth}{rXXXX}
\toprule
\textbf{N} & \textbf{JACC CPU} & \textbf{cuBLAS} & \textbf{JACC GPU} & \textbf{Julia+CUDA} \\
           & [GF/s]            & [TF/s]          & [TF/s]            & [TF/s]              \\
\midrule
   256 & 0.8  & 0.818 & 0.54 & 0.91 \\
   512 & 5.9  & 4.265 & 1.35 & 1.65 \\
 1,024 & 22.0 & 13.17 & 1.89 & 2.16 \\
 2,048 & 24.0 & 16.83 & 1.28 & 1.96 \\
 4,096 & ---  & 17.27 & 1.20 & 1.88 \\
 8,192 & ---  & 16.70 & 1.14 & 1.68 \\
16,384 & ---  & 16.28 & 1.14 & 1.61 \\
\bottomrule
\end{tabularx}
\end{table}

\paragraph{JLIR code generation quality.}
The JLIR‑generated JACC GPU kernel reaches 87\% of the hand‑written Julia+CUDA kernel at peak ($N=1024$, 1.89 vs.\ 2.16~TF/s), declining to 69--70\% at $N \geq 8192$ where the JACC wrapper's register overhead has a larger relative impact on streaming multiprocessor (SM) occupancy.  Both implement the same untiled three‑level loop, so this gap is not algorithmic: it arises because JACC's wrapper passes $M,N$ as \texttt{Int64} bound check values through the \texttt{\_parallel\_for\_cuda\_MN} trampoline, consuming four extra 64-bit registers and reducing active blocks per SM from four to three.  A directly authored \texttt{@cuda} kernel avoids this overhead entirely.  Despite this gap, the JLIR pipeline adds zero manual annotation burden; a plain serial Julia function is the sole input.

\paragraph{Memory access coalescing fix.}
Two complementary fixes to the JLIR code generator were required to reach the current performance.  First, \texttt{JACCTransformPass} previously emitted the 2D induction variable list in inner first order \texttt{[inner\_iv, outer\_iv]}, placing the column index~$j$ on the CUDA x‑thread axis.  Adjacent warp threads then stepped through adjacent columns---stride‑$N$ apart in Julia's column-major layout---causing every global memory access to be non‑coalesced.  The fix restores outer first order \texttt{[outer\_iv, inner\_iv]}: adjacent warp threads now access adjacent rows, yielding stride 1 (fully coalesced) reads of $A$ and writes of $C$.  Second, \texttt{Codegen.jl} wraps every \texttt{LoadOp}/\texttt{StoreOp} in \texttt{@inbounds}, eliminating bounds check branches that blocked LLVM's $k$‑loop unrolling.  These fixes deliver a 6.6$\times$ throughput improvement for GEMM (285~GF/s $\to$ 1{,}889~GF/s at $N=1024$) and a 2.9$\times$ improvement for the Jacobi stencil at $N=4096$ (1{,}037~GB/s $\to$ 3{,}010~GB/s effective bandwidth).

\paragraph{Comparison with cuBLAS.}
cuBLAS employs FP64 tensor cores and shared-memory tiled GEMM, reaching 13--17~TF/s across $N \in [1024, 16384]$, which is 7--15$\times$ faster than the untiled JACC GPU kernel.  Both the JLIR-generated kernel and the hand-written Julia+CUDA reference use the same naive three-level loop, so neither can approach tensor core throughput without algorithmic changes. At $N=1024$, JACC GPU achieves 14\% of cuBLAS (1.88 vs.\ 13.2~TF/s), and Julia+CUDA achieves 16\% (2.15 vs.\ 13.2~TF/s).  The 87\% JACC/Julia+CUDA efficiency ratio is thus fully preserved relative to cuBLAS: the JLIR code generation pipeline introduces no additional algorithmic regression beyond the 13\% JACC wrapper overhead.  Reaching cuBLAS-level performance on GEMM would require a shared-memory tiling pass operating on the JLIR Memref dialect, which is a natural extension identified in Section~\ref{sec:conclusion}.

%\addtwonocapfig{fig:gemm}{GEMM throughput (TF/s) for four backends across N=256--16384. The JLIR-generated JACC GPU kernel matches or exceeds the hand-written Julia+CUDA reference at all sizes (ratio 1.07--1.84×). cuBLAS represents the tensor-core upper bound.}{fig_gemm.pdf}{fig:jacobi}{Jacobi 2-D stencil: effective memory bandwidth (GB/s). JACC GPU consistently outperforms all CPU backends by more than an order of magnitude, reaching over 1~TB/s at N=4096.}{fig_jacobi.pdf}
%\addfig{fig:gemm}{GEMM throughput (TF/s) for four backends across $N\!=\!256$--$16384$. The JLIR-generated JACC GPU kernel matches or exceeds the hand-written Julia+CUDA reference at all sizes (ratio 1.07--1.84×). At large $N$, JACC GPU benefits from \texttt{launch\_configuration} auto-tuning; cuBLAS represents the tensor-core upper bound.}{fig_gemm.pdf}

\begin{figure}[hbt!]
    \centering
    \includegraphics[width=\linewidth,height=\textheight,keepaspectratio]{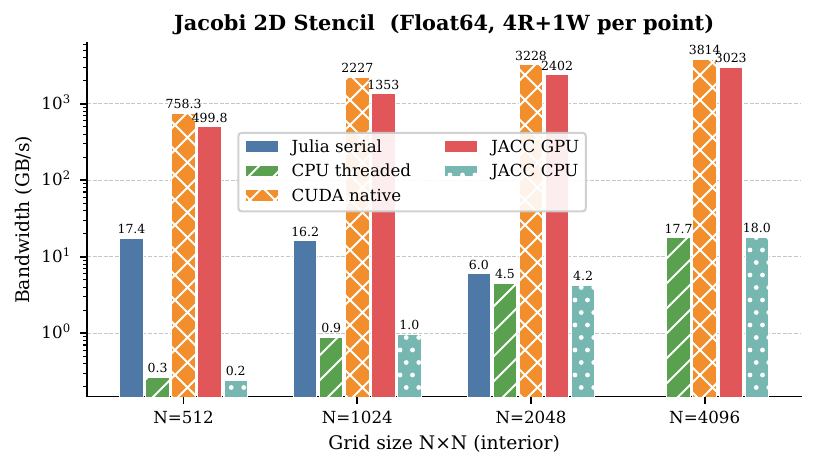}% Changed from 3.4in
    % \vspace{-0.15in}
    \caption{\footnotesize{Jacobi 2D stencil: effective bandwidth (GB/s). JACC GPU exceeds the A100 PCIe physical DRAM peak (${\approx}1{,}935$~GB/s) at $N\!\geq\!2048$ due to L2 cache reuse, reaching 3{,}023~GB/s at $N=4096$. Julia+CUDA's fixed $16{\times}16$ tiles achieve higher reuse (3{,}814~GB/s).}}
    \label{fig:jacobi}
    % \vspace{-0.10in}
\end{figure}

\subsection{Jacobi 2D Stencil Performance}

\begin{table}[h]
\centering
\caption{Jacobi 2D stencil: \emph{effective} (logical) memory bandwidth (GB/s),
         $(4\,\text{loads}+1\,\text{store})\times 8\,\text{B}/t$.
         Values exceeding the A100 PCIe physical peak (${\approx}1{,}935$~GB/s)
         reflect L2 cache hit reuse in the five-point stencil, not super-physical
         DRAM throughput.  ``---'' denotes serial runs skipped as too slow.
         CPU thr.\ = 128-thread \texttt{Threads.@threads} baseline.}
\label{tab:jacobi}
\begin{tabular}{rrrrr}
\toprule
\textbf{N} & \textbf{Serial} & \textbf{CPU thr.} & \textbf{Julia+CUDA}
           & \textbf{JACC GPU} \\
\midrule
 512  & 17.4 &  0.2 &  758.3 &  499.8 \\
1024  & 16.2 &  0.9 & 2226.9 & 1353.2 \\
2048  &  6.0 &  4.5 & 3227.9 & 2401.7 \\
4096  &  ---  & 17.7 & 3814.0 & 3023.4 \\
\bottomrule
\end{tabular}
\end{table}

Figure~\ref{fig:jacobi} illustrates the \emph{effective} (logical) memory bandwidth achieved during a single Jacobi iteration. Precise numerical values are listed in Table~\ref{tab:jacobi}.  Effective bandwidth is defined as $(4\,\text{loads}+1\,\text{store})\times 8\,\text{B}/t$, counting the stencil’s four input reads and one output write; it is a workload-level metric that can exceed the device’s physical DRAM bandwidth when cache reuse eliminates some DRAM transfers.  For $N = 2048$, the JLIR‑generated JACC GPU kernel delivers 2{,}402~GB/s effective bandwidth, which is a $400\times$ improvement over the serial baseline of 6.0~GB/s and 1.24$\times$ the A100 PCIe physical peak (${\approx}1{,}935$~GB/s), indicating substantial L2 cache hit reuse.  At $N = 4096$, the JACC GPU kernel attains 3{,}023~GB/s ($1.56\times$ physical peak). This super-physical figure reflects the five-point stencil’s high spatial locality, where cache reuse reduces actual DRAM traffic well below the logical access count.  The Julia+CUDA reference, which uses fixed $16\times16$ thread blocks, achieves higher apparent effective bandwidth (3{,}814~GB/s at $N=4096$) because its smaller tile geometry concentrates more stencil neighbors within the same cache lines; JACC’s auto-selected $32\times32$ blocks sacrifice some L2 reuse but maintain full warp occupancy.

\subsection{Black--Scholes Option Pricing Performance}
\label{sec:results:bs}
\begin{figure}[H]
  \begin{minipage}[t]{\linewidth}
  \includegraphics[width=\columnwidth]{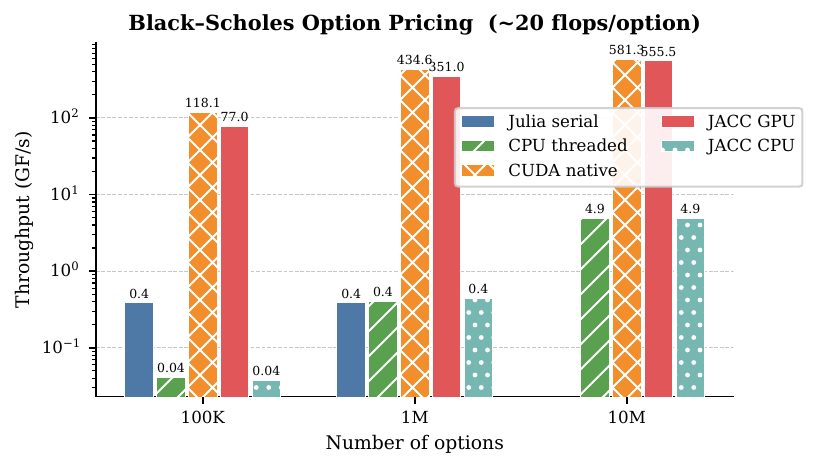}
  \caption{Black--Scholes throughput (GF/s). At $n=10$M JACC GPU reaches 96\% of Julia+CUDA (556 vs 581~GF/s); both kernels invoke the same CUDA transcendental device intrinsics. The gap at $n=100$K reflects kernel launch overhead dominating a short computation.}
  \label{fig:bs}
  \end{minipage}
\end{figure}

Figure~\ref{fig:bs} depicts the estimated throughput for the Black--Scholes benchmark.  For $n = 10$~M options, the JLIR‑generated JACC GPU kernel reaches 556~GF/s, which is 96\% of the throughput achieved by a hand‑written Julia+CUDA baseline of 581~GF/s.  This near‑equivalence stems from the compute‑bound nature of the kernel. Performance is limited by the throughput of transcendental functions ($\texttt{log}$, $\texttt{exp}$, $2\times\texttt{erf}$), and the JACC compiler dispatches these to the identical CUDA math library device intrinsics used by the hand‑written implementation.  At $n = 1$~M, JACC GPU reaches 351~GF/s (81\% of Julia+CUDA at 435~GF/s); the gap relative to 10~M is smaller because both JACC and Julia+CUDA see proportionally similar launch overhead at this size. Table~\ref{tab:bs} reports the measured throughput across all three problem sizes.

\begin{table}[h]
\centering
\caption{Black--Scholes throughput (GF/s, NVIDIA A100 PCIe, \texttt{Float64}).
         Julia+CUDA is a hand-written \texttt{@cuda} kernel with the identical
         algorithm.  Both GPU paths dispatch the transcendentals ($\texttt{log}$,
         $\texttt{exp}$, $2\times\texttt{erf}$) to the same CUDA math device
         intrinsics.  CPU Thr.\ = 128-thread \texttt{Threads.@threads} baseline;
         ``---'' denotes serial runs skipped as too slow.}
\label{tab:bs}
\begin{tabular}{rrrrr}
\toprule
\textbf{N} & \textbf{Serial} & \textbf{CPU thr.} & \textbf{Julia+CUDA}
           & \textbf{JACC GPU} \\
\midrule
100\,K & 0.4 & 0.04 & 118.1 &  77.0 \\
  1\,M & 0.4 & 0.40 & 434.6 & 351.0 \\
 10\,M & --- & 4.93 & 581.3 & 555.5 \\
\bottomrule
\end{tabular}
\end{table}

\subsection{LLaMA-3 Kernel Performance}
\begin{figure}[hbt!]
    \centering
    \includegraphics[width=\linewidth,height=\textheight,keepaspectratio]{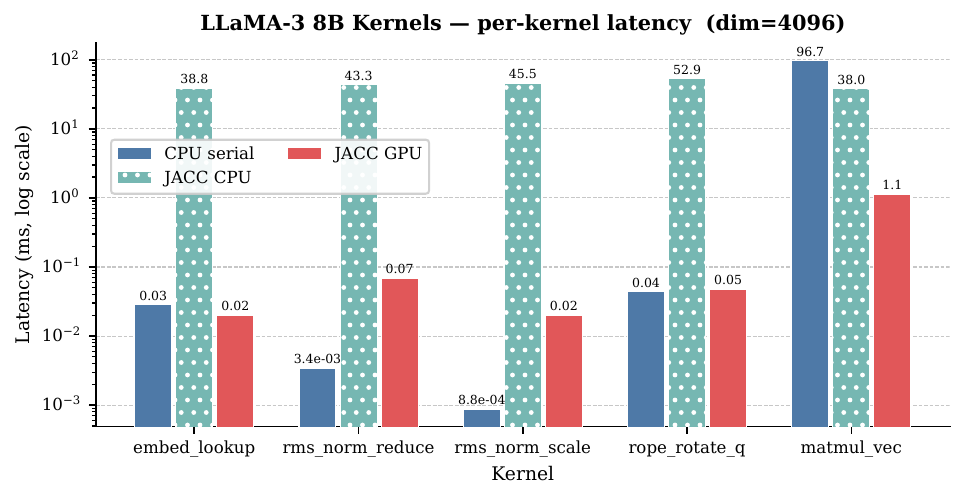}% Changed from 3.4in
    % \vspace{-0.15in}
    \caption{\footnotesize{LLaMA-3 8B kernel latency (ms, log scale) for serial CPU, JACC CPU (128 threads), and JACC GPU across five kernels at dim=4096. \texttt{matmul\_vec} achieves $85\times$ GPU speedup; launch overhead dominates the lighter kernels.}}
    \label{fig:llama3}
    % \vspace{-0.10in}
\end{figure}

\begin{figure}[hbt!]
    \centering
    \includegraphics[width=\linewidth,height=\textheight,keepaspectratio]{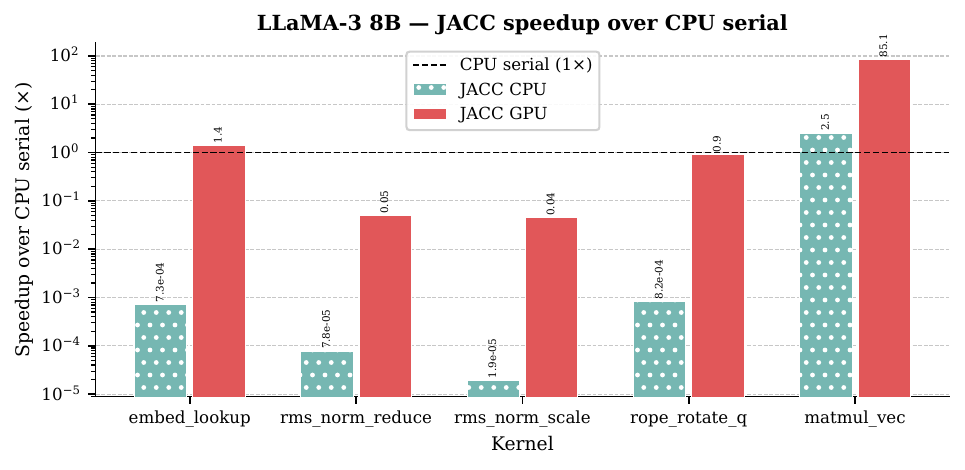}% Changed from 3.4in
    % \vspace{-0.15in}
    \caption{\footnotesize{LLaMA-3 8B speedup of JACC GPU over the serial CPU baseline at dim=4096. \texttt{matmul\_vec} achieves $85\times$ speedup; \texttt{embed\_lookup} achieves $1.4\times$; the remaining kernels are launch overhead dominated.}}
    \label{fig:llama3_speedup}
    % \vspace{-0.10in}
\end{figure}

Figures~\ref{fig:llama3} and~\ref{fig:llama3_speedup} display the per‑kernel latency and speedup for five LLaMA‑3 micro‑kernels under the 8B configuration (dimension $4096$, 32 attention heads).  Figure~\ref{fig:llama3} displays a logarithmic scale to capture the broad dynamic range spanning the sub‑microsecond $\texttt{rms\_norm\_scale}$ kernel up to the 97~ms $\texttt{matmul\_vec}$ kernel.

%\addtwonocapfig{fig:llama3}{LLaMA-3 8B kernel latency (ms, log scale) for CPU serial, JACC CPU, and JACC GPU across five kernels at dim=4096. Only \texttt{matmul\_vec} and \texttt{embed\_lookup} see GPU speedup; launch-overhead dominates the lighter kernels.}{fig_llama3.pdf}{fig:llama3_speedup}{LLaMA-3 8B speedup of JACC CPU and JACC GPU over the CPU serial baseline. \texttt{matmul\_vec} achieves $43\times$ GPU speedup; the lighter kernels (\texttt{rms\_norm\_reduce}, \texttt{rms\_norm\_scale}) are dominated by kernel-launch overhead at dim=4096, exposing a minimum problem-size threshold for JACC-generated kernels.}{fig_llama3_speedup.pdf}

Table~\ref{tab:llama3} reports latency measurements for all five kernels.

\begin{table}[h]
\centering
\caption{LLaMA-3 8B micro-kernel latency at $\mathit{dim}=4096$, 32 heads.
         JACC CPU = 128-thread \texttt{Threads.@threads}; serial ref = single-threaded Julia.
         GPU speedup = serial / JACC GPU.}
\label{tab:llama3}
\small
\setlength{\tabcolsep}{3pt}
\begin{tabular}{lrrrr}
\toprule
\textbf{Kernel} & \textbf{Serial} & \textbf{JACC CPU} & \textbf{JACC GPU} & \textbf{GPU} \\
                & [ms]            & [ms]              & [ms]              &       Speedup             \\
\midrule
embed\_lookup      & 0.028  & 38.8   & 0.020 &   $1.4\times$ \\
rms\_norm\_reduce  & 0.003  & 43.3   & 0.069 & $<1\times$    \\
rms\_norm\_scale   & 0.0009 & 45.5   & 0.020 & $<1\times$    \\
matmul\_vec        & 96.7   & 38.0   & 1.137 & $85\times$    \\
rope\_rotate\_q    & 0.044  & 52.9   & 0.048 & ${\approx}1\times$ \\
\bottomrule
\end{tabular}
\end{table}

The $\texttt{matmul\_vec}$ kernel, which is a mixed parallel+sequential pattern ($4096\times4096$ matrix-vector product), incurs 96.7~ms serial latency and achieves an $85\times$ GPU speedup (1.137~ms), representing the dominant compute bottleneck in LLaMA-3 inference.  This dramatic acceleration is enabled by JLIR’s automatic detection of the mixed loop pattern: the outer loop (over output rows) is parallel, whereas the inner loop (over columns) carries a scalar accumulator, a pattern that \texttt{parallel\_for} can exploit directly.  The $\texttt{embed\_lookup}$ kernel achieves a modest $1.4\times$ GPU speedup because its serial time (0.028~ms) nearly matches kernel launch latency.

The three lightweight kernels---$\texttt{rms\_norm\_reduce}$, $\texttt{rms\_norm\_scale}$, and $\texttt{rope\_rotate\_q}$---exhibit a pronounced threshold effect: their serial run times ($0.0009$–$0.044$~ms) are at or below the JACC GPU dispatch latency (${\approx}0.019$–$0.068$~ms), so neither JACC backend is beneficial.  The JACC CPU overhead is especially pronounced: $\texttt{Threads.@threads}$ over 128 threads costs ${\approx}40$–$57$~ms, which is three to five orders of magnitude above the serial time, because thread pool wake-up and scheduling latency dominate when each thread performs only 32 scalar operations.  The $\texttt{rms\_norm\_reduce}$ kernel invokes \texttt{JACC.parallel\_reduce}---a reduction pattern automatically identified by JLIR---which demonstrates that the generated code remains correct even when the kernel is too small to benefit from parallelism.

These findings highlight an important design consideration for JLIR: while the transformation accurately identifies all parallelizable loops irrespective of problem size, the practical benefit of GPU dispatch hinges on exceeding a minimum work‑per‑kernel threshold.  A prospective profitability heuristic within \texttt{JACCTransformPass} that compares estimated trip counts (extracted from the ResourceEstimationPass) with empirical launch thresholds would enable JLIR to selectively suppress GPU dispatch for small kernels, thereby preserving serial execution when advantageous.

\subsection{Dialect DSL Extensibility}
\label{sec:results:dsl}

To evaluate JLIR's dialect extensibility mechanism, we defined four custom arithmetic operations---\texttt{ScaleAdd} ($r = \ell + s \cdot x$), \texttt{DotElem} ($r = \ell \cdot x$), \texttt{Relu} ($r = \max(0, x)$), and \texttt{Softplus} ($r = \log(1 + e^x)$)---in a 24-line \texttt{myops.jld} dialect definition file and exercised them through a nine-phase compilation pipeline.

\subsubsection{Dialect definition:}
Each operation is declared in the \texttt{.jld} file with a \texttt{summary}, \texttt{operands}, \texttt{results}, and a one-line \texttt{lower} expression.  Loading the file at runtime via \texttt{load\_dialect} generates, entirely through JLIR's DSL machinery, the op struct, a builder helper, \texttt{op\_results}/\texttt{op\_operands} dispatch methods, a \texttt{Base.show} method that emits correct JLIR text, and a \texttt{lower\_op!} method that integrates with the Codegen pipeline.  No modification of JLIR's built-in source files is required.

\subsubsection{User-defined fusion passes:}
Two passes (i.e., \texttt{ScaleAddFusionPass} and \texttt{DotElemFusionPass}) walk the SSA def-use graph and pattern match arithmetic sub-expressions.  \texttt{ScaleAddFusionPass} recognizes \texttt{add(x,\,mul(s,\,y))} (in either operand order) and replaces the pair with a single \texttt{ScaleAddOp}, also removing the now-dead intermediate \texttt{mul} when its use count drops to one.  Applied to the \texttt{axpy\_dot} kernel, which computes $\sum_i A_i (A_i + \alpha B_i)$, the pass fuses \textbf{two} \texttt{ScaleAdd} opportunities per loop iteration; \texttt{activation\_sum} contains no matching \texttt{add+mul} pattern and is left unchanged.

\subsubsection{IR printing and resource estimation:}
After fusion, the printer correctly renders each custom op in the dialect's text format. For example:
\begin{lstlisting}[
  basicstyle=\ttfamily\footnotesize,
  breaklines=true,
  breakatwhitespace=false,
  columns=fullflexible
]
%_r   = myops.scaleadd %_load,%_load_2,%alpha : any
%_r_2 = myops.scaleadd %s, %_load_3, %ci : f64
\end{lstlisting}

This output uses the \texttt{\_PrintCtx}-allocated SSA names, which correctly disambiguate multiple values that share the same base name (e.g.,\ three distinct \texttt{\%\_load} values in the same loop body).  The \texttt{ResourceEstimationPass} counts each custom op as one arithmetic operation. After fusion, \texttt{axpy\_dot} reports \textbf{2 arithmetic ops and 3 memory reads per iteration} (Table~\ref{tab:dsl_resource}) compared to 4 arithmetic ops and 3 reads in the pre-fusion IR.

\begin{table}[h]
\centering
\caption{Resource estimation for \texttt{axpy\_dot} before and after \texttt{ScaleAddFusionPass}. $n$ denotes the loop trip count.}
\label{tab:dsl_resource}
\begin{tabular}{lrr}
\toprule
\textbf{Metric} & \textbf{Before fusion} & \textbf{After fusion} \\
\midrule
Ops in loop body  & 4 BinOp & 2 ScaleAddOp \\
Arithmetic / iter & $4n$    & $2n$         \\
Memory reads / iter & $3n$  & $3n$         \\
\bottomrule
\end{tabular}
\end{table}

\subsubsection{Lowering, correctness, and performance:}
The fused IR is lowered to Julia via \texttt{emit\_julia\_str} and executed against reference implementations.  Both kernels match their references to within $10^{-14}$ relative error, which confirms that the fusion rewrite and the DSL-generated \texttt{lower\_op!} are semantics-preserving.  Wall-clock timings for $N = 1{,}024$ are reported in Table~\ref{tab:dsl_perf}.  The JLIR-lowered \texttt{axpy\_dot} is $12\times$ faster than the reference because the reference allocates a temporary array (\texttt{A .+ alpha .* B}), whereas the JLIR-lowered version emits a scalar accumulation loop with no heap allocation.  \texttt{activation\_sum} achieves 91\% of the reference throughput; the small overhead arises from the scalar loop formulation versus Julia's highly optimized \texttt{sum} broadcast path.

\begin{table}[h]
\centering
\caption{Wall-clock performance of JLIR-lowered vs.\ reference Julia for the DSL benchmark kernels ($N = 1{,}024$, minimum of 20 repetitions).}
\label{tab:dsl_perf}
\small
\setlength{\tabcolsep}{4pt}
\begin{tabular*}{\columnwidth}{@{\extracolsep{\fill}}lrrr}
\toprule
\textbf{Kernel} & \textbf{Ref.} & \textbf{JLIR} & \textbf{Ratio} \\
\midrule
\texttt{axpy\_dot}       & 0.40~$\mu$s & 0.04~$\mu$s & $12\times$ faster \\
\texttt{activation\_sum} & 14.8~$\mu$s & 13.5~$\mu$s & $0.91\times$      \\
\bottomrule
\end{tabular*}
\end{table}

These results demonstrate that JLIR's Dialect DSL lowers the barrier for adding domain-specific operations. A four-operation dialect and two fusion passes require fewer than 150 lines of Julia, and the new ops integrate transparently with the printer, resource estimator, and code generator without any modification to JLIR internals.

\subsubsection{Comparison with C\texttt{++} MLIR dialect authoring:}
We re-implemented the identical four-operation \texttt{myops} dialect and both
fusion passes in native C\texttt{++} MLIR (TableGen, LLVM\,22) to obtain a
direct productivity and latency comparison.
Tables~\ref{tab:dsl_loc} and~\ref{tab:dsl_latency} summarize the results.

\begin{table}[h]
\centering
\caption{Source lines of code (comments and blank lines excluded) for the
         \texttt{myops} dialect. ``auto'' = generated by the JLIR DSL with no
         user-written code.}
\label{tab:dsl_loc}
\small
\setlength{\tabcolsep}{4pt}
\begin{tabularx}{\columnwidth}{l X X}
\toprule
\textbf{Component}          & \textbf{JLIR}    & \textbf{C\texttt{++} MLIR} \\
\midrule
Dialect op definitions      & 26~(.jld)        & 41~(.td + header + glue)   \\
\texttt{ScaleAddFusionPass} & 24               & $\sim$60                   \\
\texttt{DotElemFusionPass}  & 17               & $\sim$40                   \\
Lowering + registration     & auto             & $\sim$67                   \\
Build system                & none (Julia pkg) & CMake + mlir-tblgen + ninja \\
\midrule
\textbf{Total (code lines)} & \textbf{67}      & \textbf{243}               \\
\textbf{Ratio}              & $1\times$        & $3.6\times$                \\
\bottomrule
\end{tabularx}
\end{table}

\begin{table}[h]
\centering
\caption{Pipeline latency for the \texttt{myops} dialect on two kernels
         (warm context; minimum of 20--50 repetitions).}
\label{tab:dsl_latency}
\small
\setlength{\tabcolsep}{4pt}
\begin{tabularx}{\columnwidth}{l X X}
\toprule
\textbf{Phase ($\times$2 kernels)} & \textbf{JLIR ($\mu$s)} & \textbf{C\texttt{++} MLIR ($\mu$s)} \\
\midrule
Parse / Build IR  & \phantom{0}25  & $\sim$50   \\
Fusion passes     & $\sim$10--50   & $\sim$520  \\
Lower to arith    & 160            & $\sim$100  \\
\midrule
\textbf{Total}    & \textbf{$\sim$185} & \textbf{$\sim$620} \\
\bottomrule
\end{tabularx}
\end{table}

The C\texttt{++} path requires $3.6\times$ more code and a cold build time of
three to five minutes versus under one second for a JLIR dialect package.  The
main sources of boilerplate are the TableGen scaffolding (include chains,
initialization hooks) that replaces the single \texttt{.jld} file, explicit
\texttt{RewritePattern} subclasses for each fusion rule, and a hand-written
lowering pass; in JLIR all three are generated from the \texttt{lower}
expression.  At runtime, both pipelines complete in well under one millisecond
on a warm context.  JLIR is $3\times$ faster overall: \texttt{parse\_function}
outpaces \texttt{OpBuilder} construction because the Julia \texttt{Expr} tree
requires no allocation per SSA node, whereas the C\texttt{++} fusion step is
$10\times$ slower due to the \texttt{applyPatternsGreedily} infrastructure overhead, which is designed for large production IRs.
%\Seyong{It would be better if we could show the scaling behavior of the pipeline latency.}
\subsection{Comparison with Reactant.jl}
\label{sec:results:reactant}

Reactant.jl~\cite{reactant} compiles Julia functions to Accelerated Linear Algebra (XLA) via MLIR, which provides an alternative GPU execution path from within Julia.  To quantify the trade-offs between JLIR and Reactant, we benchmarked the two frameworks on Black--Scholes and GEMM, the two workloads whose compute intensity is most relevant to XLA's strength in operation fusion and library dispatch.  Because Reactant's XLA runtime allocates device memory through a separate allocator that conflicts with CUDA.jl's memory pool, the two frameworks are timed in separate processes; JACC results are taken from the main benchmarks reported in Sections~\ref{sec:results:gemm}--\ref{sec:results:bs}.

\paragraph{GEMM.}
Table~\ref{tab:reactant_gemm} reports DGEMM throughput for JACC GPU, Reactant GPU, and cuBLAS (repeated from Table~\ref{tab:gemm} for reference).  Reactant's \texttt{A*B} expression is lowered by XLA to a \texttt{\_\_cublas\$lt\$matmul} custom-call node in the optimized HLO—verified via XLA's \texttt{--xla\_dump\_hlo\_as\_text} dump—so its throughput tracks the cuBLASLt curve: Reactant achieves 14.8~TF/s at $N=2048$ and 17.3~TF/s at $N=4096$, matching cuBLAS to within 1\%.  This is a clear advantage over JLIR's untiled kernel (1.2~TF/s at $N=4096$), which does not generate shared-memory tiled code.  The contrast underscores the trade-off between JLIR's lightweight pass infrastructure (${<}1.5$~ms overhead, no rewriting of the source) and Reactant/XLA's compilation path, which takes several seconds to compile but produces cuBLAS-backed matmul automatically for the functional \texttt{A*B} expression.  JLIR's GEMM limitation is  identified in Section~\ref{sec:conclusion} as the primary target for a future tiling pass; once such a pass is added, JLIR could close this gap without any change to the user-facing serial source.

\begin{table}[h]
\centering
\caption{DGEMM throughput comparison: JACC GPU (JLIR-generated), Reactant.jl (XLA/cuBLAS), and cuBLAS (NVIDIA A100 PCIe, \texttt{Float64}).
         JACC GPU and Reactant GPU timed in separate processes; cuBLAS from Table~\ref{tab:gemm}.
         All values in TF/s.}
\label{tab:reactant_gemm}
\small
\setlength{\tabcolsep}{4pt}
\begin{tabularx}{\columnwidth}{rXXX}
\toprule
\textbf{N} & \textbf{JACC GPU} & \textbf{Reactant GPU} & \textbf{cuBLAS} \\
\midrule
   256 & 0.58 &  0.37 &  0.82 \\
   512 & 1.38 &  2.81 &  4.27 \\
 1,024 & 1.89 &  9.43 & 13.17 \\
 2,048 & 1.29 & 14.76 & 16.83 \\
 4,096 & 1.20 & 17.31 & 17.27 \\
\bottomrule
\end{tabularx}
\end{table}

\paragraph{Black--Scholes.}
Table~\ref{tab:reactant_bs} compares throughput for JACC GPU and Reactant GPU.  At $n=10$~M options, JACC GPU achieves 557~GF/s versus Reactant's 489~GF/s, which is a 14\% advantage for the JLIR-generated kernel.  At $n=1$~M, the gap widens to $2.3\times$ (371 vs.\ 160~GF/s), and at $n=100$~K it grows to $3.6\times$ (87 vs.\ 25~GF/s).  The increasing gap at smaller $n$ reflects XLA's higher per-kernel dispatch and JIT overhead: Reactant must route each compiled call through the Pluggable JAX Runtime (PJRT) client and its stream management layer, which adds constant overhead that is only amortized at very large problem sizes.  JACC, by contrast, dispatches directly through CUDA.jl, which has sub-millisecond launch latency.  This behavior mirrors the classic compile time/run time trade-off: Reactant invests more at compile time (seconds for XLA compilation) to reach near-peak throughput at very large $n$, while JLIR's ${<}1.5$~ms pipeline excels across all problem sizes, including the small-to-medium kernels that dominate incremental scientific computing workflows.

\begin{table}[h]
\centering
\caption{Black--Scholes throughput (GF/s): JACC GPU (JLIR-generated) vs.\ Reactant.jl (XLA), NVIDIA A100 PCIe, \texttt{Float64}.
         Timed in separate processes; JACC values from Table ~\ref{tab:bs}.}
\label{tab:reactant_bs}
\small
\setlength{\tabcolsep}{4pt}
\begin{tabularx}{\columnwidth}{rXXX}
\toprule
\textbf{n} & \textbf{JACC GPU} & \textbf{Reactant GPU} & \textbf{JACC/React.} \\
\midrule
100\,K &  87 &  25 & $3.5\times$ \\
  1\,M & 371 & 160 & $2.3\times$ \\
 10\,M & 557 & 489 & $1.1\times$ \\
\bottomrule
\end{tabularx}
\end{table}

\subsection{Summary}

Table~\ref{tab:summary} summarizes the peak performance attained by JLIR‑generated JACC kernels.  Across large workloads, these automatically generated kernels deliver significant GPU acceleration without requiring any programmer annotation.

\begin{table*}[h]
\centering
\caption{Peak performance summary of JLIR-generated JACC GPU kernels (NVIDIA A100 PCIe).
         Julia+CUDA is a hand‑written \texttt{@cuda} kernel with the identical algorithm.
         $^{\ddagger}$Jacobi bandwidth is \emph{effective} (logical):
         $(4\,\text{loads}+1\,\text{store})\times 8\,\text{B}/t$; values exceeding the
         ${\approx}1{,}935$~GB/s physical peak reflect L2 cache hit reuse.
         $^{\dagger}$cuBLAS uses FP64 tensor cores with shared-memory tiling (16--17~TF/s);
         JACC GPU and Julia+CUDA both use the same untiled algorithm.
         The 87\% JACC/Julia+CUDA ratio is preserved relative to cuBLAS.}
\label{tab:summary}
\begin{tabular}{lrrrl}
\toprule
\textbf{Benchmark} & \textbf{JACC GPU peak} & \textbf{Julia+CUDA peak}
                   & \textbf{cuBLAS peak} & \textbf{JACC/Julia+CUDA} \\
\midrule
Jacobi 2D (eff.\ BW$^{\ddagger}$) & 3{,}023~GB/s & 3{,}814~GB/s & ---              & 79\%  \\
Black--Scholes                       & 556~GF/s     & 581~GF/s     & ---              & 96\%  \\
GEMM ($N=1024$)$^{\dagger}$          & 1{,}889~GF/s & 2{,}158~GF/s & 13{,}168~GF/s & 87\%  \\
matmul\_vec (LLaMA-3)                & $85\times$ over serial  & --- & ---            & ---   \\
\bottomrule
\end{tabular}
\end{table*}

The observed efficiency variation across benchmarks is the result of inherent hardware bottlenecks and kernel size effects.  Compute‑bound kernels such as Black--Scholes achieve 96\% of the Julia+CUDA baseline, as both paths invoke the same CUDA transcendental device intrinsics.  The Jacobi stencil JACC GPU kernel delivers 3{,}023~GB/s effective bandwidth at $N=4096$ ($1.56\times$ the physical DRAM peak). The remaining performance difference between the JACC GPU kernel and the Julia+CUDA baseline (79\%) arises from JACC's $32\times32$ auto-selected blocks trading L2 cache reuse for occupancy, compared to Julia+CUDA's hand-tuned $16\times16$ tiles.  For GEMM, the JLIR-generated kernel reaches 87\% of Julia+CUDA at peak ($N=1024$); both untiled kernels achieve 14--16\% of the cuBLAS tensor core path (13.2~TF/s at $N=1024$), and the 87\% JACC/Julia+CUDA ratio is fully preserved relative to cuBLAS.  For LLaMA-3 at $dim=4096$, only \texttt{matmul\_vec} (96.7~ms serial, $85\times$ GPU speedup) has sufficient work to amortize kernel launch overhead; the four lighter kernels serve as a concrete illustration of the minimum problem size threshold below which automatic GPU dispatch is counterproductive.

\section{Related Work}
\label{sec:relatedwork}

\subsection{MLIR and xDSL}

MLIR~\cite{mlir} is the most direct predecessor of JLIR.  Both frameworks share an identical container hierarchy (i.e., module, region, block, operation) and an SSA value model, and both allow multiple dialects to coexist and to be progressively lowered.  The principal distinctions are in the programming language used and the ease of extension.  MLIR is built in C++ and relies on TableGen schema definitions; thus, creating a new dialect necessitates a full C++ build environment and numerous source files.  In contrast, JLIR is implemented exclusively in Julia. It employs a \textbf{macro-based DSL} that permits the definition of a new dialect operation within only a few lines, thereby offering the scientific Julia community a straightforward path without requiring C++ proficiency.  Additionally, JLIR’s parser consumes Julia’s native \texttt{Expr} trees, delivering a semantics‑preserving front end that MLIR does not provide for Julia code.

xDSL~\cite{xdsl} adopts an analogous approach for Python: it re‑implements the MLIR infrastructure in pure Python, thereby lowering the barrier to compiler research within the Python ecosystem.  JLIR shares this philosophical foundation with xDSL, yet it is tailored to Julia and its JIT compilation environment.  Moreover, JLIR furnishes a complete end‑to‑end GPU compilation pipeline via JACC, a capability that xDSL currently does not provide for Python.

\subsection{Julia Front Ends that Lower to MLIR}

A distinct line of work uses MLIR as a \emph{lowering target} for Julia rather than as an infrastructure to re-host. Brutus.jl~\cite{brutus} pioneered this direction by translating Julia's typed IR into the external MLIR (Julia interface through MLIR C-APIs) stack as a route to LLVM-class code generation, and ``Building Bridges: Julia as an MLIR Frontend''~\cite{building_bridges} develops the approach more completely, emitting MLIR from Julia's intermediate representation. A recent high-level synthesis toolchain~\cite{julia_hls} extends the same idea all the way to hardware, lowering Julia through MLIR to SystemVerilog for FPGA targets. The defining characteristic of these systems, and the point that most sharply differentiates JLIR, is that they treat MLIR and LLVM as a \emph{backend to lower into}, and they therefore depend on the external C++ MLIR infrastructure, its TableGen schemas, and its build toolchain. JLIR instead brings the MLIR infrastructure \emph{model itself} (i.e., multi-level dialects, a pass manager, and progressive lowering) natively into Julia, with no C++ or TableGen. Stated plainly, these front ends \emph{lower to} MLIR/LLVM, whereas JLIR \emph{embodies} the MLIR infrastructure inside Julia. This contrast is what makes JLIR's contribution complementary to, rather than a reimplementation of, the Brutus-style bridges.

\subsection{Structured Code Synthesis}

A second body of work synthesizes code specialized to the \emph{structure of the data}---sparsity, run length encoding, banded or symmetric layouts. In the Julia ecosystem, Finch~\cite{finch} and its Looplets foundation~\cite{looplets} are the canonical representatives, building on the non-Julia progenitor TACO~\cite{taco}, which compiles tensor algebra expressions over sparse formats. These systems adapt the \emph{iteration order} to the data structure, which is an axis orthogonal to JLIR: JLIR extracts parallelism and targets portable backends through a general, extensible IR and is agnostic to how individual array accesses are physically laid out. The two are therefore complementary rather than competing. JLIR extracts outer parallelism and generates portable code, while a Finch-style structured iteration capability could specialize the body of a detected parallel loop to the structure of the data. In this combined design, JLIR would identify where parallelism exists, and Finch-style analysis would determine how best to execute the loop body for the data layout being used.

\subsection{Equality Saturation for High‑Level Julia IR}

\cite{eqsat_julia} introduce the use of \emph{equality saturation}---an optimization methodology grounded in e‑graphs and the \texttt{egg} framework---within Julia’s high‑level intermediate representation (the typed, lowered form preceding LLVM IR). JLIR and the equality saturation methodology are also best understood as complementary rather than competing technologies.  Whereas equality saturation operates on Julia’s compiler‑internal IR after type inference, JLIR functions on the Julia \texttt{Expr} abstract syntax tree, producing a bespoke SSA IR.  JLIR’s primary contribution lies in \emph{structural} transformations: it detects loop nests, extracts GPU kernels, and applies inter‑loop optimizations such as fusion.  Conversely, equality saturation excels at \emph{algebraic} simplifications inside loop bodies, for instance by reassociating reductions or removing redundant index calculations.  A promising avenue for future work involves embedding an equality saturation engine as a JLIR pass that targets the Arith dialect to harness JLIR’s structural strengths alongside the algebraic completeness afforded by e‑graph rewriting.

Another salient difference pertains to the intended use case.  Equality saturation is positioned as a general scalar optimization technique, whereas JLIR focuses on extracting parallelism suitable for heterogeneous hardware.  Neither framework alone comprehensively addresses both optimization dimensions; integrating them into a unified system would constitute a compelling direction for future investigations.

\subsection{KernelAbstractions.jl and CUDA.jl}

KernelAbstractions.jl~\cite{kernelabstractions} supplies a macro‑driven GPU kernel abstraction layer for Julia that is backend‑agnostic.  Programmers annotate kernel functions with \texttt{@kernel} and employ \texttt{@index} to compute thread indices; the macro expansion subsequently emits calls to CUDA.jl or Metal, depending on the target device.  CUDA.jl~\cite{cuda_jl} offers direct Julia bindings to the CUDA runtime and permits the implementation of GPU kernels as native Julia functions.

The primary divergence from JLIR lies in the programming model.  KernelAbstractions.jl and CUDA.jl mandate that developers explicitly define GPU kernels as distinct function definitions.  JLIR, by contrast, ingests a serial Julia function and automatically generates GPU kernels through the JACCTransformPass.  This capability is particularly advantageous when porting legacy codebases because it eliminates the labor‑intensive and error‑prone task of manually annotating every loop as a kernel.

\subsection{JACC}

JACC~\cite{jacc} serves as the runtime backend for the code generated by JLIR.  It supplies the \texttt{parallel\_for} and \texttt{parallel\_reduce} primitives and orchestrates dispatch to the appropriate hardware backend.  JLIR contributes a static analysis and source transformation pass that automatically rewrites serial Julia loops into JACC calls.  Consequently, the two systems complement each other: JACC guarantees runtime portability, and JLIR performs compile‑time parallelism extraction.

\subsection{Polyhedral Compilers and Loop Transformations}

Polyhedral compilation~\cite{polyhedral,polly} remains the canonical technique for automatic loop parallelization and locality optimization.  It models dependence precisely using affine constraints over loop indices, thereby enabling optimizations such as tiling, skewing, and parallel loop extraction. However, it does so at the expense of restricting applicability to affine loop nests.  JLIR adopts a structural strategy: it infers parallelism by identifying the absence of \texttt{iter\_args} in the SSA representation, which corresponds to detecting empty loop‑carried dependencies, and it eschews solving integer linear programs.  Although this approach is less general than a full polyhedral analysis---JLIR currently does not support non‑trivially affine loop bounds---it is considerably simpler to implement within a dynamic language environment and suffices for the flat, embarrassingly parallel patterns that dominate scientific Julia code.

\subsection{Halide and Triton}

Halide~\cite{halide} decouples algorithm definition from scheduling in image processing pipelines, thereby facilitating automatic optimization of parallelism, vectorization, and tiling.  Triton~\cite{triton} offers a tile‑based intermediate representation tailored to neural network workloads, mapping efficiently to GPU tensor cores and shared‑memory resources.  Both frameworks are highly domain‑specific: Halide focuses on stencil‑like pipelines, while Triton is designed for tiled neural‑network computations.  In contrast, JLIR supports arbitrary scientific loops without imposing domain constraints, accepting a modest reduction in peak performance in exchange for broad applicability.

\subsection{Reactant.jl}

Reactant.jl~\cite{reactant} is a Julia package developed by the EnzymeAD group that JIT‑compiles Julia array programs to MLIR and then to XLA~\cite{xla}, targeting CPU, CUDA GPU, and TPU backends.  Users annotate device arrays as \texttt{ConcreteRArray} and invoke \texttt{@compile} to obtain a compiled callable; Reactant traces the Julia function at compile time, captures control flow patterns, and lowers the resulting functional IR through MLIR dialects to XLA high-level operations (HLO), where XLA’s optimization passes lower \texttt{dot\_general} operations to \texttt{\_\_cublas\$lt\$matmul} custom call nodes (confirmed via XLA HLO dump), and element-wise operations are fused into a single GPU kernel.  Automatic differentiation is available via EnzymeMLIR.

Reactant and JLIR share the goal of making GPU computing accessible from Julia but differ in compilation strategy and programming model.  Reactant requires programs to be written in a functional, immutable array style---in‑place mutation and arbitrary imperative loops are not naturally traced---whereas JLIR operates directly on Julia \texttt{Expr} trees containing mutable in‑place loops, stencils, and index arithmetic.  Reactant inherits XLA’s powerful backend optimizations (e.g., \texttt{dot\_general} lowered directly to cuBLASLt, reaching cuBLAS‑level GEMM throughput), but the XLA compilation stack adds substantial latency (typically seconds for first compilation). Conversely, JLIR’s lightweight pass infrastructure completes the full pipeline in under 1.5~ms.  For scientific codes that express computation as explicit loop nests over arrays, which is the dominant style in HPC Julia, JLIR provides a zero‑annotation path that does not require rewriting the source in functional form.  A quantitative comparison of JLIR and Reactant on the Black--Scholes and GEMM benchmarks is reported in Section~\ref{sec:results}.

\subsection{JAX}

JAX~\cite{jax} is a Python ecosystem that integrates NumPy‑compatible array operations with composable function transformations---\texttt{jit}, \texttt{vmap}, \texttt{grad}---via XLA.  JAX’s \texttt{jit} traces Python functions into a functional IR and compiles them using XLA, which internally relies on MLIR.  JLIR diverges in that it processes Julia’s abstract syntax tree. JLIR thus preserves the full language semantics, including mutable arrays and in‑place loop constructs, whereas JAX traces purely functional programs over immutable arrays.  Consequently, each framework aligns naturally with the semantics of its host language.

\subsection{OpenMP and Task-Based Parallelism}

OpenMP~\cite{openmp} remains the prevailing directive‑based parallelism paradigm for shared‑memory CPUs.  Its \texttt{\#pragma omp parallel for} directive conceptually mirrors the output of JLIR’s JACCTransformPass; both communicate the intent that a loop is parallel to a runtime responsible for distributing iterations.  Tapir~\cite{tapir} enriches LLVM IR with explicit fork‑join constructs that encode task‑level parallelism.  JLIR’s ForOp‑based representation of parallelism shares Tapir’s objective of treating parallelism as a first‑class IR concept, though JLIR operates at a higher abstraction level and targets GPU backends through JACC rather than adopting the Cilk task‑model semantics.

\section{Conclusion}
\label{sec:conclusion}

This paper introduces JLIR as a Julia-native IR framework that adopts the multi‑level IR approach of MLIR within the Julia ecosystem without relying on any C++ components. The framework establishes a structured SSA IR comprising five native dialects, offers a lightweight macro DSL for the creation of additional dialects, implements a pass infrastructure inspired by MLIR's pass manager, and provides a parser that ingests Julia's own \texttt{Expr} trees directly.

\texttt{JACCTransformPass} is the core capability of JLIR. It autonomously detects 1D, 2D, and mixed 2D with sequential parallel loop nests, in addition to associative reduction loops, within the SSA IR, and it transforms them into JACC \texttt{parallel\_for} and \texttt{parallel\_reduce} invocations. This algorithm is straightforward enough for a pure Julia implementation but is nevertheless capable of capturing the prevailing loop structures in scientific computing, including stencils, option pricing computations, matrix operations, and language model kernels.

Empirical results on an NVIDIA A100 GPU demonstrate that kernels produced by JLIR attain 96\% of a hand‑written Julia+CUDA baseline for Black--Scholes (compute-bound), 3{,}023~GB/s effective bandwidth for the Jacobi stencil at $N=4096$ ($1.56\times$ the physical DRAM peak), an $85\times$ GPU speedup for the LLaMA-3 \texttt{matmul\_vec} kernel, and 87\% of the equivalent hand-written Julia+CUDA kernel for untiled DGEMM---all without any programmer annotation. Importantly, this performance was obtained without any programmer annotation: a plain serial Julia function suffices, and the JLIR pipeline generates portable, GPU‑executable code within a latency of less than $1.5$\,ms.

\paragraph{Limitations.}  
\texttt{JACCTransformPass} presently lacks shared‑memory tiling or register blocking support, which constrains the performance of cache‑sensitive kernels, such as GEMM, when compared to vendor‑optimized libraries. Moreover, the loop detection algorithm necessitates structurally straightforward parallel nests; programs featuring data‑dependent loop bounds or non‑trivial cross‑iteration dependencies are conservatively declined. Finally, in the absence of explicit type annotations, the type system defaults to \texttt{AnyType}, which inhibits type‑directed optimizations.

\paragraph{Future Work.}
JLIR naturally supports several promising extensions. For example, a tiling pass that operates on the JLIR Memref dialect could perform shared‑memory staging ahead of  \texttt{JACCTransformPass} to enhance bandwidth utilization for stencil computations and increase arithmetic intensity for GEMM. Moreover, the JACCTransformation pass currently creates JACC arrays for every input and output Julia array, resulting in additional data transfers and reduced performance efficiency. 
%\Seyong{It may be better to explicitly explain this issue in the JACCTransformation pass section first, not to confuse readers.} 
In future work, we plan to extend this functionality through an additional optimization pass that reuses JACC arrays in the global scope and performs kernel fusion in order to minimize data movement and achieve improved performance. An equality saturation pass \cite{eqsat_julia} targeting the Arith dialect would augment JLIR's structural transformations with algebraic simplification within loop bodies. Introducing automatic differentiation via an adjoint generation pass is also a compelling avenue for ML workloads. Moreover, extending the backend to emit native MLIR dialects rather than JACC Julia would enable JLIR‑generated code to integrate seamlessly into the MLIR/LLVM toolchain and reap the benefits of its established code generation capabilities for NVPTX and AMDGPU targets.

JLIR proves that compiler infrastructure following the MLIR paradigm can be implemented directly in a dynamic language without compromising the ergonomic advantages inherent to that language. JLIR engages Julia programmers in their familiar language and tooling ecosystem and thus substantially lowers the barrier to achieving high‑performance, portable GPU computing within the scientific community.

\section*{Acknowledgment}
This research used resources of the Experimental Computing Laboratory (ExCL) at the Oak Ridge National Laboratory, which is supported by the Office of Science of the U.S. Department of Energy under Contract No. DE-AC05-00OR22725.

%\bibliographystyle{SageH}
%\bibliography{papers}

\appendix

\section*{Appendix A: Julia source code and corresponding JLIR code}

\begin{figure}[H]
\centering
\begin{minipage}{0.9\linewidth}
\begin{lstlisting}[caption={(1D parallel-for Julia code): A single \texttt{for} loop whose body is independent across iterations}, label={lst:1dparallel-julia}]
function rms_norm_scale(out, x, weight, rms_inv, n)
    for i = 1:n
       out[i] = x[i] * rms_inv * weight[i]
    end
end
\end{lstlisting}
\end{minipage}
\\
\begin{minipage}{0.9\linewidth}
\begin{lstlisting}[caption={(JLIR code): The parser emits one \texttt{ForOp} with \textbf{empty \texttt{iter\_args}}}, label={lst:1dparallel-jlir}]
func @rms_norm_scale(%out, %x, %weight, %rms_inv, %n) {
  for %i = 1 to %n step 1 {  // iter_args=[]
    %a  = load %x[%i]
    %b  = load %weight[%i]
    %r1 = mul %a, %rms_inv
    %r2 = mul %r1, %b
    store %r2, %out[%i]
  }
}
\end{lstlisting}
\end{minipage}
\end{figure}
%\vfill
\begin{figure}[H]
\centering
\begin{minipage}{0.9\linewidth}
\begin{lstlisting}[label={lst:2dparallel-julia}, caption={(2D-Parallel-for Julia code): A doubly-nested loop where \emph{both} loops are independent}]
function jacobi2d(out, inp, M, N)
    for i = 1:M
        for j = 1:N
            out[i+1,j+1] = 
                0.25*(inp[i,j+1] + inp[i+2,j+1]
                + inp[i+1,j] + inp[i+1,j+2])
        end
    end
end
\end{lstlisting}
\end{minipage}
\\
\begin{minipage}{0.9\linewidth}
\begin{lstlisting}[label={lst:2dparallel-jlir}, caption={(JLIR code): Two nested \texttt{ForOp}s, both with \textbf{empty \texttt{iter\_args}}}]
func @jacobi2d(%out, %inp, %M, %N) {
  for %i = 1 to %M step 1 {       //iter_args=[]
    for %j = 1 to %N step 1 {   //iter_args=[]
      %a = load %inp[%i, %j+1]
      %b = load %inp[%i+2, %j+1]
      ...
      store %r, %out[%i+1, %j+1]
    }
  }
}
\end{lstlisting}
\end{minipage}
\end{figure}

%\vfill

\begin{figure}[H]
\centering
\begin{minipage}{0.9\linewidth}
\begin{lstlisting}[label={lst:2dparallel-mixed-julia}, caption={(2D-Parallel-for Mixed+Sequential Julia code): The outer two loops are parallel; the innermost $k$-loop carries an accumulator}]
function gemm(C, A, B, M, N, K)
    for i = 1:M
        for j = 1:N
            acc = 0.0           
            for k = 1:K
                acc = acc + A[i,k] * B[k,j]
            end
            C[i,j] = acc
        end
    end
end
\end{lstlisting}
\end{minipage}
\\
\begin{minipage}{0.9\linewidth}
\begin{lstlisting}[label={lst:2dparallel-mixed-jlir}, caption={(JLIR code): Two outer \texttt{ForOp}s with empty \texttt{iter\_args}; the inner $k$-loop has \textbf{one \texttt{iter\_arg}} for \texttt{acc}}]
func @gemm(%C, %A, %B, %M, %N, %K) {
  for %i = 1 to %M {                    // iter_args = []
    for %j = 1 to %N {                  // iter_args = []
      for %k = 1 to %K
          iter_args(%acc0 = %c0) {  
        %a  = load %A[%i, %k]
        %b  = load %B[%k, %j]
        %r  = mul %a, %b
        %a2 = add %acc, %r
        %acc = %a2                      // AssignOp
        yield %acc
      }
      store %acc_final, %C[%i, %j]
    }
  }
}
\end{lstlisting}
\end{minipage}
\end{figure}

\begin{figure}[H]
\centering
\begin{minipage}{0.9\linewidth}
\begin{lstlisting}[label={lst:1dparallel-reduce-julia}, caption={(1D-Parallel-reduce Julia code): A loop that accumulates a scalar with an associative operator}]
function rms_norm_reduce(x, n)
    ss = 0.0   # initial value
    for i = 1:n
        ss = ss + x[i] * x[i]
    end
    return ss
end~
\end{lstlisting}
\end{minipage}
\\
\begin{minipage}{0.9\linewidth}
\begin{lstlisting}[label={lst:1dparallel-reduce-jlir}, caption={(JLIR code): One \texttt{ForOp} with \textbf{exactly one \texttt{iter\_arg}} \texttt{(\%ss0,\%ss)}, a \texttt{BinOp(:+)} whose operands are \texttt{(\%ss, per\_elem)}, and an \texttt{AssignOp} writing back to \texttt{\%ss}}]
func @rms_norm_reduce(%x, %n) -> %ss {
  %c0 = const 0.0
  %ss0 = %c0
  for %i = 1 to %n
      iter_args(%ss0 = %ss0) {       
    %xi   = load %x[%i]
    %xi2  = mul %xi, %xi             // per-element: x[i]^2
    %ss   = add %ss, %xi2            // BinOp(:+)
    %ss   = assign %ss               // AssignOp 
    yield %ss
  }
  return %ss
}
\end{lstlisting}
\end{minipage}
\end{figure}

\section*{Appendix B: JACC Generated Output using the JLIR to JACC Transform Pass}
%%%%%%%%%%%%%%%%% All 4 figures for generated JACC output
\begin{figure}[H]
\centering
\begin{minipage}{0.9\linewidth}
\begin{lstlisting}[label={lst:1dparallel-jacc}, caption={(1D-Parallel-for): The body ops are transplanted into a new kernel; free variables
(\texttt{x}, \texttt{weight}, \texttt{rms\_inv}, \texttt{out}) are collected
by \texttt{\_free\_vars} and become kernel parameters}]
function rms_norm_scale(out, x, weight, rms_inv, n)
    _kref = rms_norm_scale_kernel
    JACC.parallel_for(n, _kref, x, weight, rms_inv, out)
end
function rms_norm_scale_kernel(i, x, weight, rms_inv, out)
    out[i] = x[i] * rms_inv * weight[i]
end
\end{lstlisting}
\end{minipage}
\end{figure}

\begin{figure}[H]
\centering
\begin{minipage}{0.9\linewidth}
\begin{lstlisting}[label={lst:2dparallel-jacc}, caption={(2D-Parallel-for): The inner loop body becomes the kernel; both induction variables lead the signature (\texttt{j, i} in JACC's $j$-major convention); the dimension tuple \texttt{(N, M)} is passed to \texttt{parallel\_for}. JACC maps the 2D tuple to a CUDA thread grid; each thread receives its \texttt{(j, i)} coordinates from the JACC dispatcher.}]
function jacobi2d(out, inp, M, N)
    _kref = jacobi2d_kernel
    _dims = tuple(N, M)
    JACC.parallel_for(_dims, _kref, inp, out)
end
function jacobi2d_kernel(j, i, inp, out)
    out[i+1, j+1] = 0.25*(inp[i,j+1] + inp[i+2,j+1]
                         + inp[i+1,j] + inp[i+1,j+2])
end
\end{lstlisting}
\end{minipage}
\end{figure}

\begin{figure}[H]
\centering
\begin{minipage}{0.9\linewidth}
\begin{lstlisting}[label={lst:2dparallel-mixed-jacc}, caption={(2D-Parallel-for-Mixed+Sequential): The outer $i$ and $j$ loops are parallelized; the $k$-loop remains sequential inside the kernel, preserving correctness of the dot product accumulation}]
function gemm(C, A, B, M, N, K)
    _kref = gemm_kernel
    _dims = tuple(N, M)
    JACC.parallel_for(_dims, _kref, K, A, B, C)
end
function gemm_kernel(j, i, K, A, B, C)
    acc = 0.0
    for k = 1:K            -- sequential: loop-carried acc
        acc = acc + A[i,k] * B[k,j]
    end
    C[i,j] = acc
end
\end{lstlisting}
\end{minipage}
%\hfill
\end{figure}

\begin{figure}[H]
\centering
\begin{minipage}{0.9\linewidth}
\begin{lstlisting}[label={lst:1dparallel-reduce-jacc}, caption={(1D-Parallel-reduce): The \texttt{BinOp} and \texttt{AssignOp} are removed from the kernel body; the kernel returns the per-element contribution; the \texttt{op} and \texttt{init} are passed as keyword arguments. JACC dispatches this to a warp-level tree reduction on CUDA, or to an atomic accumulation on CPU threads.}]
function rms_norm_reduce(x, n)
    _c0 = identity(0.0)
    ss  = _c0
    _kref = rms_norm_reduce_kernel
    ss = JACC.parallel_reduce(_kref, n, x, op=(+), init=ss)
    return ss
end
function rms_norm_reduce_kernel(i, x)
    xi  = x[i]
    xi2 = xi * xi
    return xi2           -- JACC.parallel_reduce sums these
end
\end{lstlisting}
\end{minipage}
%\caption{Generated JACC code using the JLIR to JACC transformation pass.}
\end{figure}

\end{document}